\documentclass[aps,pra,twocolumn,a4paper,showpacs,superscriptaddress,floatfix,10pt]{revtex4}
\usepackage[pdftex]{graphicx,color}
\usepackage{dcolumn}
\usepackage{bm}

\usepackage{amsfonts}
\usepackage{amssymb}
\usepackage{amsmath}
\usepackage{amsthm}
\usepackage{xcolor}

\usepackage{subfigure}
\usepackage{rotate}
\usepackage{dsfont}
\usepackage{hyperref}
\usepackage{breakurl}
\usepackage{multirow}
\usepackage{bbm}

\usepackage{pdfsync}

\let\oldmarginpar\marginpar
\renewcommand\marginpar[1]{\-\oldmarginpar[\raggedleft\tiny #1]%
{\raggedright\tiny #1}}

\def\KeyWord#1{$\backslash$\IfColor{$\!\!$\textRed{#1}\textBlack}{#1}$\!\!$}

\newcommand{\be}{\begin{equation} }
\newcommand{\ee}{\end{equation} }
\newcommand{\ba}{\begin{eqnarray} }
\newcommand{\ea}{\end{eqnarray} }

\newcommand{\bit}{\begin{itemize}}
\newcommand{\eit}{\end{itemize}}

\DeclareFontFamily{U}{mathx}{\hyphenchar\font45}
\DeclareFontShape{U}{mathx}{m}{n}{
      <5> <6> <7> <8> <9> <10>
      <10.95> <12> <14.4> <17.28> <20.74> <24.88>
      mathx10
      }{}
\DeclareSymbolFont{mathx}{U}{mathx}{m}{n}
\DeclareFontSubstitution{U}{mathx}{m}{n}
\DeclareMathAccent{\widecheck}{0}{mathx}{"71}
\DeclareMathAccent{\wideparen}{0}{mathx}{"75}

\renewcommand\[{\begin{equation}}
\renewcommand\]{\end{equation}}

\graphicspath{{FinalFigures/}}

\begin{document}
\title{Weak-Coupling Theory of Pair Density-Wave Instabilities in Transition Metal Dichalcogenides}

\author{Daniel Shaffer}

\affiliation
{
Department  of  Physics,  Emory  University,  400 Dowman Drive, Atlanta,  GA  30322,  USA
}
\affiliation
{
School of Physics and Astronomy, University of Minnesota, Minneapolis, MN 55455, USA
}

\author{F. J. Burnell}

\affiliation
{
School of Physics and Astronomy, University of Minnesota, Minneapolis, MN 55455, USA
}

\author{Rafael M. Fernandes}

\affiliation
{
School of Physics and Astronomy, University of Minnesota, Minneapolis, MN 55455, USA
}

\begin{abstract}
The possibility of realizing pair density wave (PDW) phases, in which Cooper pairs have a finite momentum, presents an interesting challenge that has been studied in a wide variety of systems.  In conventional superconductors, this is only possible when external fields lift the spin degeneracy of the Fermi surface, leading to pair formation at an incommensurate momentum.  Here, we study a second possibility, potentially relevant to transition metal dichalcogenides, in which the Fermi surface consists of a pair of pockets centered at the $\pm K$ points of the Brillouin zone as well as a central pocket at the $\Gamma$ point. In the limit where these three pockets are identical, the pairing susceptibility has a logarithmic divergence at the non-zero wave-vectors $\pm \mathbf{K}$, allowing for a weak-coupling analysis of the PDW instability. We find that repulsive electronic interactions combine to yield effective attractive interactions in the singlet and triplet PDW channels, as long as the $\Gamma$ pocket is present. Because these PDW channels decouple from the uniform superconducting channel, they can become the leading unconventional pairing instability of the system. Upon solving the linearized gap equations, we find that the PDW instability is robust against small trigonal warping of the $\pm K$ pockets and small detuning between the $\Gamma$ and $\pm K$ pockets, which affect the PDW transition in a similar way as the Zeeman magnetic field affects the uniform superconducting transition. We also derive the Ginzburg-Landau free energy for the PDW gaps with momenta $\pm \mathbf{K}$, analyzing the conditions for and consequences of the emergence of FF-type and LO-type PDW ground states. Our classification of the induced orders in each ground state reveals unusual phases, including an odd-frequency charge-$2e$ superconductor in the LO-type PDW.
\end{abstract}

\maketitle

\section{Introduction}

A pair density wave (PDW) is an exotic quantum state of matter in which the amplitude and/or phase of the superconducting (SC) gap function display real-space periodicity \cite{FF,LO,Himeda02,Berg07,Aperis08,Agterberg08, AgterbergSigrist09, BergFradkinKivelson09, Agterberg11, Lee14, Casalbuoni04, HaimKhodas21, SchradeFu21} -- in other words, the Cooper pairs have finite momentum (for a recent review, see \cite{AgterbergRev20}). Experimentally, signatures consistent with a PDW order have been reported in cuprates \cite{HamidianDavis16, Edkins19},  NbSe$_2$ \cite{LiuDavis20}, iron-based superconductors \cite{Liu22}, kagome superconductors \cite{Chen21}, and UTe$_2$ \cite{AishwaryaFradkinMadhavan22,GuPaglioneDavis22}. Theoretically, it is well established that a Pauli-limited BCS-like superconductor can acquire a non-zero modulation upon application of a sufficiently large magnetic field \cite{FF,LO}. In non-centrosymmetric crystals, a small field applied along certain directions is enough to induce such a modulation \cite{BarzykinGorkov02, AgterbergKaur07}. A secondary periodic gap also emerges when a uniform SC state microscopically coexists with another ordered state that breaks the translational symmetry of the lattice, such as spin or charge density waves \cite{PsaltakisFenton83,MurakamiFukuyama98,Kyung00,Aperis08,AlmeidaFernandes17}.

In all these cases, however, the pairing instability of the system is towards a uniform SC state, i.e. the pairing susceptibility $\chi_{\mathrm{SC}}(\mathbf{Q})$ diverges for $\mathbf{Q} = 0$. It is the breaking of another symmetry of the system (time-reversal, inversion, or translational) by other degrees of freedom that induces a finite modulation in the gap function. In contrast, models in which the leading pairing instability is towards a PDW phase are much rarer \cite{AgterbergRev20}. From a weak-coupling perspective, the issue stems from the fact that the non-interacting pairing susceptibility does not generally display a log-divergence except for $\mathbf{Q} = 0$. To overcome this difficulty, several microscopic models with moderate or strong interactions have been proposed to stabilize a PDW ground state \cite{Zachar01,BergFradkinKivelson10,Loder10,JaefariFradkin12,SotoGarridoFradkin15,Wardh18,HanKivelson20,SlagleFu20,WuRaghu22,SettyHirschfeldPhillipsYang22}. A PDW phase was also reported to emerge within a spin-fermion model \cite{WangAgterbergChubukovPRB15} and an Amperean pairing model \cite{Lee14}. Moreover, extensive numerical simulations have shown that the doped Hubbard model also displays PDW-like correlations \cite{Mai22}.

Another class of systems that can potentially support a PDW state are those whose band dispersions $\epsilon(\mathbf{p})$ display particle-particle nesting for a finite wave-vector $\mathbf{Q}$, i.e. $\epsilon(-\mathbf{p}) = \epsilon(\mathbf{p}+\mathbf{Q})$ (this is not to be confused with the particle-hole nesting condition $\epsilon(-\mathbf{p}) = -\epsilon(\mathbf{p}+\mathbf{Q})$, which is relevant for spin density waves). In this case, the pairing susceptibility has a logarithmic divergence also at the nesting wave-vector, which opens up the possibility of a weak-coupling PDW instability. Physically, the nesting condition is satisfied, for example, if two circular Fermi pockets are centered at  $\pm \mathbf{Q}/2$ and $\mathbf{Q}$ is not a reciprocal lattice vector. Such a situation can occur, for instance, in doped Weyl semimetals -- although time-reversal and inversion symmetries are often explicitly broken in these systems \cite{ChoMoore12,BednikBurkov15}.   Interestingly, a metal in the electronic nematic spin-triplet state also has a pairing susceptibility peaked at finite momentum \cite{SotoGarridoFradkin14}.

Nearly identical Fermi pockets can also emerge in the honeycomb or triangular lattices at the $\pm K$ points of the Brillouin zone -- for instance, in doped graphene \cite{LeeKim18}. As discussed in Ref. \cite{KimNatCom17}, the band structures of several doped transition metal dichalcogenides (TMDs), such as MoS$_2$ and NbSe$_2$, display Fermi pockets at $\pm K$ (see Fig. \ref{FigFS}), making these materials potential candidates for a PDW instability -- see also     \cite{TsuchiyaSigrist16, VenderleyEunAhKim19}. More broadly, superconducting TMDs, particularly in monolayer form, have been widely investigated in the past decade due to the unique interplay between pairing and the Ising spin-orbit coupling that emerges from the breaking of the inversion symmetry by the monolayer \cite{Exp1TMD,Exp1MoS2,Exp2MoS2,Exp3MoS2,Exp4MoS2,LawPRL14,MakNat16,LawMay16,LawMay16II,Oiwa18,Oiwa19,Khodas18,Shaffer20,HamillPribiag20,Lortz20,WickramaratneAgterbergMazin20,MargalitBerg21}.  

While the presence of nearly-circular $\pm \mathrm{K}$ Fermi pockets ensures a peak of the pairing susceptibility at the wave-vectors $\pm \mathbf{K}$, a PDW instability requires interactions that are attractive in the PDW channel. While phonons generate attractive pairing interactions, they are expected to favor conventional uniform SC. On the other hand, electronic interactions can promote effective attractive interactions in unconventional pairing channels.  Ref. \cite{KimNatCom17} showed that, in the case of a monolayer TMD with fully spin-valley-polarized Fermi pockets at the $\pm K$ points, attraction in the PDW channel only emerges via a Kohn-Luttinger-like mechanism that is third-order in the Hubbard interaction, thus implying a very small transition temperature within a weak-coupling approach. 

In this paper, motivated by the band structure of TMDs and by these previous results, we revisit the problem of a weak-coupling PDW instability in a triangular/honeycomb lattice with Fermi pockets at $\pm K$ and at $\Gamma$. By considering all eight symmetry-allowed spin-independent repulsive electronic interactions involving the low-energy electronic states, we find that the presence of the $\Gamma$ pocket is essential to stabilize a PDW instability that can outcompete the uniform SC instability. Interestingly, the $\Gamma$ pocket is present in several TMDs, such as NbSe$_2$ and sufficiently hole-doped MoS$_2$ \cite{Mattheiss73,Rossnagel01,ManzeliNature17}.

Our analysis reveals several qualitative properties of the PDW instability and of the PDW ordered state. First, when the three pockets centered at $\pm K$ and $\Gamma$ are circular and perfectly nested, the PDW and SC channels decouple at the mean-field level, with four interaction terms contributing to the uniform SC gap equations (as previously shown by us and Kang in Ref. \cite{Shaffer20}) and the other four terms contributing to the PDW gap equations. As a result, whether PDW or SC is the dominant instability is a matter of the relative strength between the microscopic interactions. In particular, we find that attraction in the PDW channel requires sufficiently large inter-pocket exchange and Umklapp interactions between the $\Gamma$ and $\pm K$ pockets, with the former (latter) favoring a triplet (singlet) PDW state.

Perfect nesting, however, is not a realistic property of TMD compounds.  To account for this, we consider the effects of both trigonal warping at the $\pm K$ pockets, as well as of different Fermi momenta at $\pm K$ and $\Gamma$. Both of these perturbations tune the system away from the perfect nesting condition of three identical circular Fermi pockets.  We find that when PDW is the dominant instability at perfect nesting, it remains robust over a finite range of these detuning parameters, which impact the PDW state in the same way that a Zeeman field impacts a uniform s-wave SC state.

In our linearized gap equations, we find that two distinct PDW gaps can condense, with inequivalent wave-vectors $\mathbf{K}$ and $-\mathbf{K}$. To elucidate whether only one of them condenses, resulting in a time-reversal symmetry-breaking FF-type (Fulde-Farrell) PDW, or whether both condense with the same gap amplitude, implying a time-reversal symmetry preserving LO-type (Larkin-Ovchinikov) PDW, we derive and analyze the Ginzburg-Landau free energy up to sixth-order terms.  We find that both types of PDW phases are possible in our model.  We further show that each of these types of PDW is associated with a distinct set of induced orders, similarly to the previously studied cases of a PDW on the square lattice and of a PDW on the honeycomb lattice with wave-vector $\mathbf{M}$ \cite{Agterberg08, BergFradkinKivelson09,WangAgterbergChubukovPRL15,WangAgterbergChubukovPRB15,BrydonAgterberg19}. Specifically, while the LO-type PDW induces a charge density wave as well as charge-$2e$, charge-$4e$ and charge-$6e$ superconductivity, the FF-type PDW induces loop current order and charge-$6e$ superconductivity. Interestingly, we find that, depending on the sign of the sixth-order term of the free energy, the induced charge-$2e$ state corresponds to either even-frequency or odd-frequency uniform superconductivity. Finally, we briefly discuss the impact to our results of the Ising SOC present in monolayer TMDs.   

The remainder of this paper is structured as follows. In Sec. \ref{Sec:Model} we present our model of the Fermi surface and discuss the generic pairing interactions between low-energy electrons. In Sec. \ref{SecNoSOC} we analyze the  linearized mean-field gap equation of our model,  determining when a PDW instability can occur.  In Sec. \ref{SecGL}, we derive and analyze the Ginzburg-Landau free energy of the PDW, which is required to fully determine the symmetry of the PDW state, and to identify possible induced orders. Finally, in Sec. \ref{Sec:Discussion} we summarize our results and briefly discuss possible effects due to spin-orbit coupling, including topological PDW phases.

\section{Interacting Microscopic Model} \label{Sec:Model}

\subsection{Single-Body Hamiltonian and Fermi Surfaces}

Our starting point is a simple model to describe layered TMDs  \cite{Mattheiss73, Rossnagel01, Xiao12, LawMay16, Shaffer20} consisting of three Fermi pockets: a $\Gamma$ pocket at the Brillouin zone center, and two $K$ pockets centered at the distinct zone corners $\pm K$.  The single-body Hamiltonian that describes these Fermi surfaces is
\[H=\sum_{\eta\mathbf{p}\alpha}\epsilon_\eta(\mathbf{p}) d^\dagger_{\mathbf{p}\eta\alpha}d_{\mathbf{p}\eta\alpha}\label{H0}\]
Here \(\eta=\Gamma,\pm K\) is a  a pocket index, and  \(\mathbf{p}\) denotes the (small) momentum as measured from the pocket center. 
Thus the annihilation operator \(d_{\mathbf{p}\eta\alpha}\) annihilates an electron with spin \(\alpha\) and momentum \(\mathbf{p}\) measured relative to the pocket center \(\eta\).   \(\epsilon_\eta (\mathbf{p}) \) is the dispersion about the pocket $\eta$, which we take to be independent of the electron's spin.  As discussed in \cite{Shaffer20}, this approximation is sufficient to determine the nature of the dominant Fermi surface instability.

\begin{figure}[htp]
	\centering
	\includegraphics[width=0.475\textwidth]{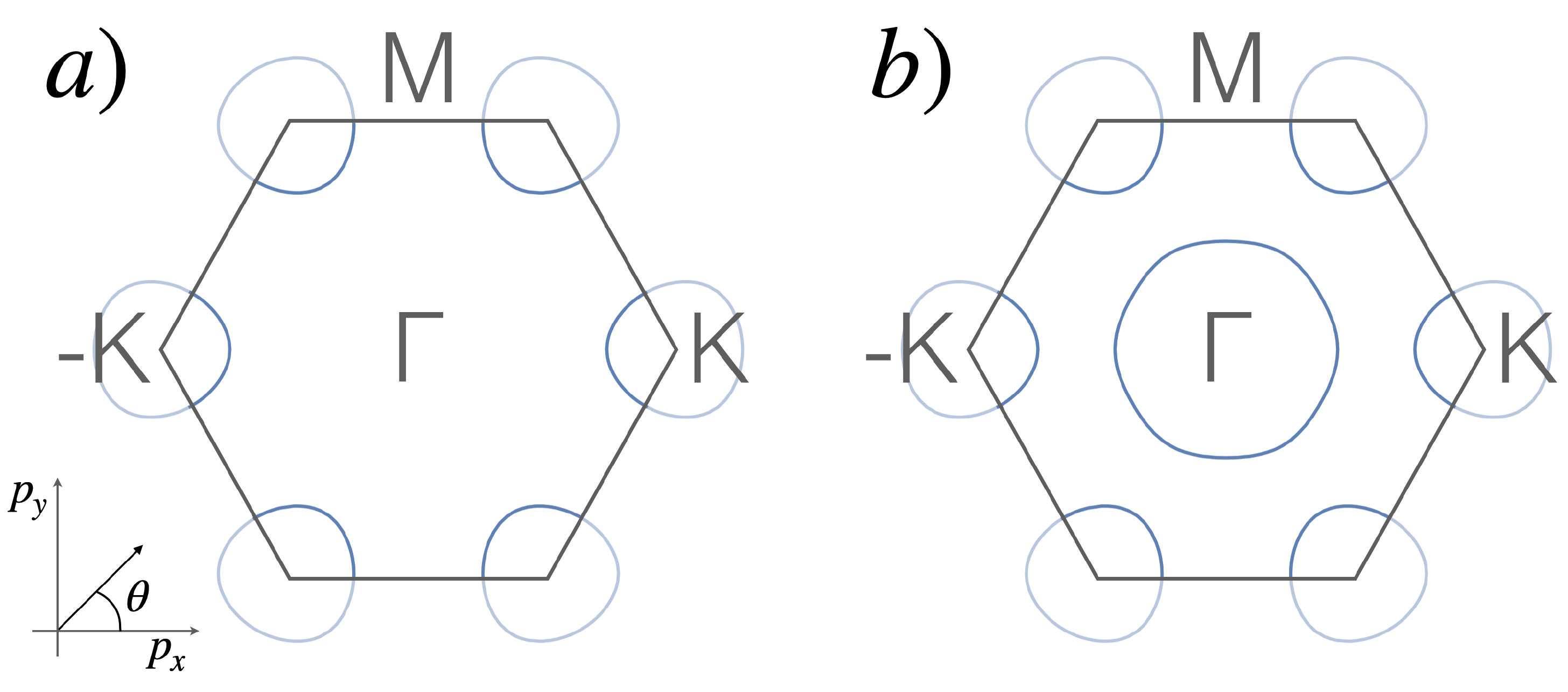}
	\caption{Fermi surfaces and the Brillouin zone of systems considered in this work without (a) and with (b) a Fermi pocket at the \(\Gamma\) point. All pockets are assumed to be hole pockets, although the results in this work apply also for systems with all electron pockets.}
	\label{FigFS}
\end{figure}

Figs. \ref{FigFS} (a) and (b) show two types of Fermi surfaces relevant to certain TMDs without spin-orbit coupling. For instance, while the central hole pocket is absent in MoS$_2$, it is present in NbSe$_2$  \cite{Mattheiss73, Rossnagel01, Xiao12}. As we discuss later, the presence of the $\Gamma$ pocket has important consequences for the emergence of a PDW state. Both dispersions are obtained from a simplified tight-binding model summarized in Appendix \ref{AppendixA}. Note that in either case, the Fermi surfaces are not perfectly circular,  exhibiting hexagonal and trigonal warping at \(\Gamma\) and \(\pm K\) points. In addition, generically the Fermi momenta of the \(\Gamma\) and \(K\) pockets are not equal.  

We will argue in the next section that the hexagonal warping of the central pocket does not qualitatively affect the pairing instabilities.  However, both the trigonal warping at \(K\) points and the Fermi momentum mismatch between $\Gamma$ and $K$ do significantly suppress the PDW instability.   We therefore take the dispersions to be
\[\epsilon_\Gamma(\mathbf{p})=-\frac{p^2}{2m_\Gamma}-\mu_\Gamma \label{epsGamma}\]
and
\[\epsilon_{\pm K}(\mathbf{p})=-\frac{p^2}{2m_K}-\mu_K\pm w\cos3\theta \label{epsK}\]
where \(\theta\) is the angle made by \(\mathbf{p}\) from the \(p_x\) axis and \(w\) parametrizes trigonal warping. Note that the chemical potential of the system is given by $\mu = (\mu_\Gamma + \mu_K)/2$ whereas $|\mu_\Gamma - \mu_K|/2$ gives the mismatch between the bottom of the central and corner bands. Inserting this expression into Eq. (\ref{H0}) gives the minimal model needed to discuss the stability of PDW phases in generic 1H TMD systems.

\subsection{Interactions Near the Fermi Surfaces}

\begin{figure*}[htp]
	\centering
	\includegraphics[width=0.8\textwidth]{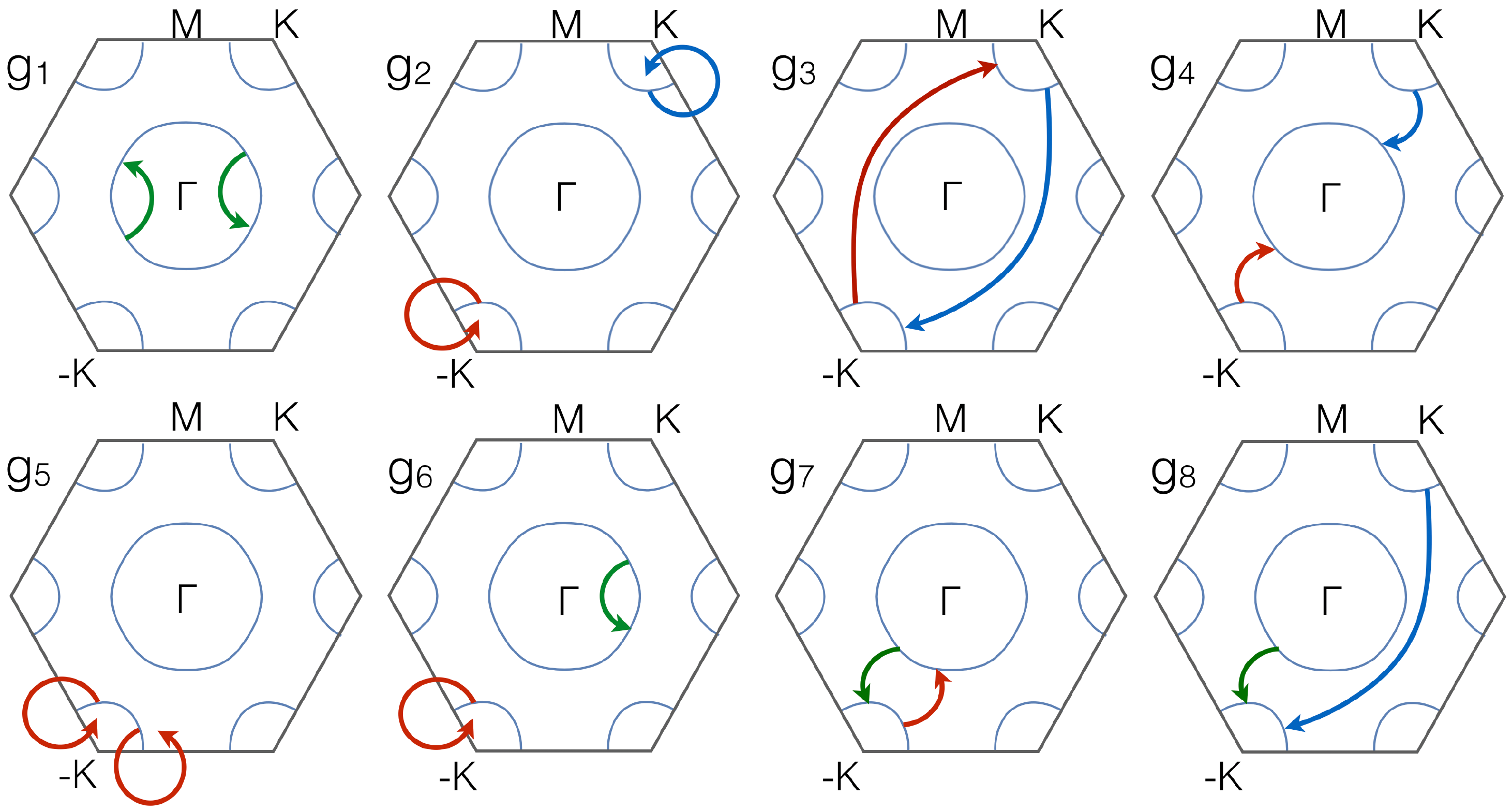}
	\caption{ \label{FigInt} Interaction channels \(g_n\) involving low-energy states near the Fermi pockets shown in the Brillouin zone. The arrows indicate which pocket the fermions start and end on before and after interacting, with the arrow's head representing an electron creation operator at a momentum ${\bf k}$ (relative to the pocket center), and the arrow's end representing an electron annihilation operator at a different momentum ${\bf p}$ (relative to the pocket center).}
\end{figure*}

To study the pairing instabilities at weak coupling, we consider generic symmetry allowed spin- and momentum-conserving interactions that are uniform along the Fermi surface, i.e. independent of the momentum \(\mathbf{p}\).  Here we will assume that the microscopic interactions are repulsive, as is the case if they  arise, for example, from density-density interactions. A similar approach was previously used in different contexts, such as iron-based superconductors (for a review on the latter, see \cite{RafaelAndrey16}). Near the Fermi surface, these interactions can be decomposed into eight channels, which we parametrize by eight momentum-independent coupling constants $g_1, \dots , g_8$.   Fig. \ref{FigInt} shows the scattering process  associated with each of these channels.    Four of them, parameterized by the couplings $g_1, \dots g_4$, mediate interactions between Cooper pairs of electrons with zero total momentum, and therefore lead to a uniform superconducting instability.  These are intrapocket density-density \(g_1\) at \(\Gamma\), interpocket density-density \(g_2\) at \(\pm K\), exchange \(g_3\) between \(\pm K\), and pair-hopping \(g_4\) between \(\Gamma\) and \(\pm K\). The role of these interactions in promoting unconventional superconducting states was previously studied in \cite{Shaffer20}.
 
\begin{figure*}[htp]
	\centering
	\includegraphics[width=0.8\textwidth]{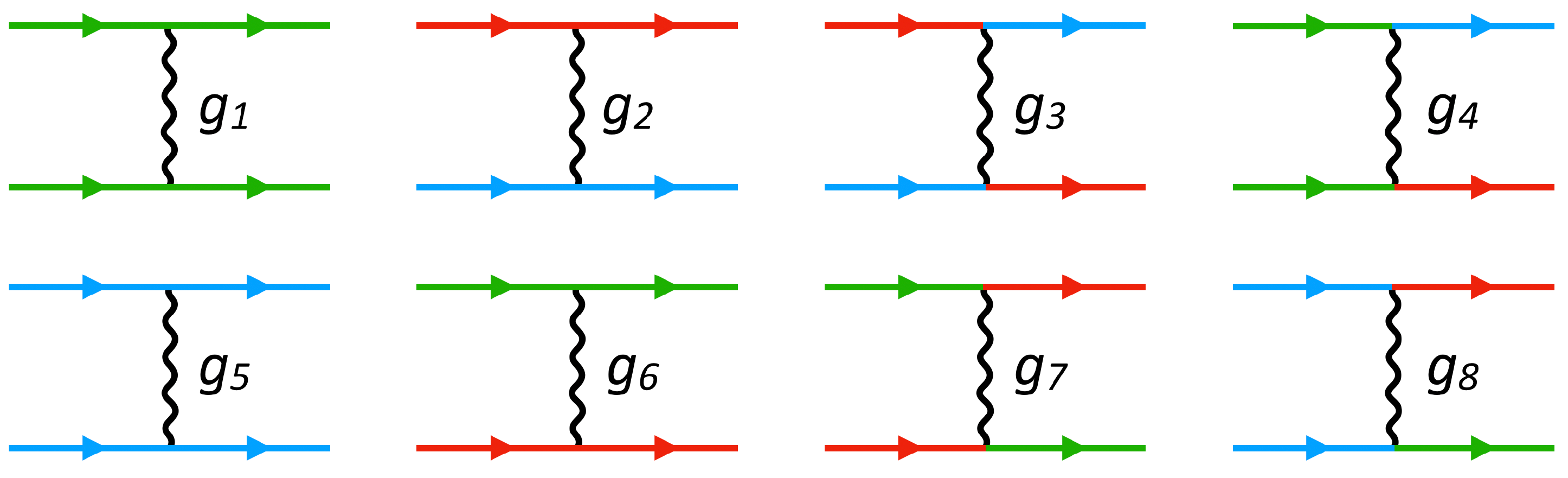}
	\caption{\label{FigIntDiag} Feynman diagrams corresponding to the eight interaction processes in Fig. \ref{FigInt}. Green, blue and red colors correspond to \(\Gamma\), \(K\) and \(-K\) pockets respectively, and spin is conserved at each vertex.  For each diagram involving a $K$-pocket fermion, there is a corresponding diagram with a fermion at $-K$; we have suppressed some of these diagrams for brevity.}
\end{figure*}

 The other four interactions, parameterized by $g_5, \dots g_8$, mediate pairing interactions between Cooper pairs with a total momentum \(\pm 2K=\mp K\):  intrapocket density-density \(g_5\) at \(\pm K\), interpocket  density-density \(g_6\) between \(\Gamma\) and \(\pm K\), exchange \(g_7\) between \(\Gamma\) and \(\pm K\), and scattering \(g_8\) from a pair at \(\Gamma\) and \(\pm K\) to a pair at \(\mp K\). The latter process is allowed by Umklapp because \(3\mathbf{K}=0\). The Feynman diagrams representing these eight processes are shown in Fig. \ref{FigIntDiag}. 

 The resulting interaction Hamiltonian has the form:
\begin{align} \label{Eq:Hint}
H_{\text{Int}}=&\frac{1}{2}\sum_{\alpha\beta, \mathbf{p} \mathbf{k}}  \big(g_1d^\dagger_{\mathbf{k}\Gamma \alpha}d^\dagger_{-\mathbf{k}\Gamma  \beta}d_{\mathbf{p}\Gamma \beta}d_{-\mathbf{p}\Gamma \alpha} \\
&+g_2 d^\dagger_{\mathbf{k}K\alpha}d^\dagger_{-\mathbf{k}-K \beta }d_{\mathbf{p}-K\beta}d_{-\mathbf{p}K \alpha} \nonumber\\
&+g_3 d^\dagger_{\mathbf{k}K\alpha}d^\dagger_{-\mathbf{k}-K \beta}d_{\mathbf{p}K\beta}d_{-\mathbf{p}-K \alpha} \nonumber\\
&+g_4 d^\dagger_{\mathbf{k}\Gamma \alpha}d^\dagger_{-\mathbf{k}\Gamma \beta }d_{\mathbf{p}K\beta}d_{-\mathbf{p}-K \alpha} \nonumber\\
&+g_5  d^\dagger_{\mathbf{k} K \alpha}d^\dagger_{-\mathbf{k} K  \beta }d_{\mathbf{p}K \beta}d_{-\mathbf{p}K  \alpha} \nonumber\\
&+g_6 d^\dagger_{\mathbf{k}\Gamma \alpha}d^\dagger_{-\mathbf{k} -K \beta }d_{\mathbf{p} -K \beta}d_{-\mathbf{p} \Gamma  \alpha} \nonumber\\
&+g_7 d^\dagger_{\mathbf{k}\Gamma \alpha}d^\dagger_{-\mathbf{k} -K \beta}d_{\mathbf{p} \Gamma \beta}d_{-\mathbf{p}  - K  \alpha} \nonumber\\
&+g_8  d^\dagger_{\mathbf{k} K \alpha}d^\dagger_{-\mathbf{k}K  \beta} d_{\mathbf{p} \Gamma \beta}d_{-\mathbf{p}-K \alpha}  \nonumber\\
&+h.c. + K \leftrightarrow - K \nonumber
\end{align}
where \(\alpha\) and \(\beta\) are spin indices. Because the microscopic  interactions are repulsive, all coupling constants are positive.

It is convenient to express the interactions (\ref{Eq:Hint}) in terms of their spin-singlet and spin-triplet components.   In the  PDW-channel, the interactions associated with $g_5$ and $g_8$ are symmetric under exchange of the pocket indices; these lead to singlet channel interactions.  The remaining two terms, associated with $g_6$ and $g_7$, are related by exchanging the pocket indices of the two annihilation operators; thus the symmetric (anti-symmetric) combination of these two leads to a singlet (triplet) channel instability.  A similar situation holds in the uniform superconducting channel \cite{Shaffer20}.  

To make this decomposition explicit, we express the interaction Hamiltonian   in the form: 
\begin{equation}
\begin{aligned}
H_{\text{Int}}= \frac{1}{2}\sum_{\substack{\mathbf{p,k}\\ \eta, \eta', \zeta, \zeta'} }\sum_{\alpha\beta\alpha'\beta'} & \left ( [V^s]^{\eta'\zeta';\alpha'\beta'}_{\eta\zeta;\alpha\beta} +[V^t]^{\eta'\zeta';\alpha'\beta'}_{\eta\zeta;\alpha\beta} \right )  \\
& \times d^\dagger_{\mathbf{k}\eta'\alpha'}d^\dagger_{-\mathbf{k}\zeta'\beta'}d_{-\mathbf{p}\zeta\beta}d_{\mathbf{p}\eta\alpha} \label{HV}
\end{aligned}
\end{equation}

The PDW channel interactions are then given by:
\begin{align} \label{Eq:Ints_SpinChannel}
\left[V^{s}\right]^{KK;\alpha'\beta'}_{KK;\alpha\beta} & =  g_{5}(i\sigma^{y})^{\alpha\beta}(i\sigma^{y})^{\alpha'\beta'}  \\
\left[V^{s}\right]^{KK;\alpha'\beta'}_{\Gamma,- K;\alpha\beta} & =  g_{8}(i\sigma^{y})^{\alpha\beta}(i\sigma^{y})^{\alpha'\beta'}\nonumber \\
\left[V^{s}\right]^{\Gamma,- K;\alpha'\beta'}_{\Gamma,- K;\alpha\beta}  & =  \frac{1}{2}(g_{6}+g_{7})(i\sigma^{y})^{\alpha\beta}(i\sigma^{y})^{\alpha'\beta'}\nonumber \\
\left[V^{t}\right]^{\Gamma,- K;\alpha'\beta'}_{\Gamma,- K;\alpha\beta} & =  \frac{1}{2}(g_{6}-g_{7})\sum_{j=x,y,z}(\sigma^{j}i\sigma^{y})_{\alpha\beta}^{*}(\sigma^{j}i\sigma^{y})^{\alpha'\beta'}\nonumber \ ,
\end{align}
The expressions in the uniform SC channel are analogous, with $g_k \rightarrow g_{k-4}$, as previously shown in \cite{Shaffer20}.

As we will argue later, we must also include the following weak momentum dependent interactions in the triplet channels in order to fully determine the gap functions in our PDW phases: 
\begin{widetext}
\begin{align} \label{Eq:IntsKdep}
\left[V^{t}(\mathbf{p};\mathbf{k})\right]^{KK;\alpha'\beta'}_{KK;\alpha\beta} & =  g_{5}^{t}\cos(3\theta_{\mathbf{k}})\cos(3\theta_{\mathbf{p}})(\sigma^{j}i\sigma^{y})_{\alpha\beta}^{*}(\sigma^{j}i\sigma^{y})^{\alpha'\beta'}\nonumber \\
\left[V^{t}(\mathbf{p};\mathbf{k})\right]^{KK;\alpha'\beta'}_{\Gamma,-K;\alpha\beta} & =  \sqrt{2}g_{8}^{t}\cos(3\theta_{\mathbf{k}})(\sigma^{j}i\sigma^{y})_{\alpha\beta}^{*}(\sigma^{j}i\sigma^{y})^{\alpha'\beta'}\nonumber \\
\end{align}
\end{widetext}
Analogously to the case of uniform SC discussed in Ref. \cite{Shaffer20}, these interactions are not needed to drive the instabilities that we describe below.  Without them, however, some symmetry-allowed components of the gap functions vanish. In this work, we will consider $g_{5}^t$ and $g_8^t$ small corrections compared to the interactions that drive the superconducting instabilities of the system.

Note that although we do not explicitly include them in our model, isotropic spin fluctuations are expected to mainly change the values of the coupling constants in the singlet/triplet channels. Moreover, anisotropic spin-fluctuations, which naturally emerge in the presence of spin-orbit coupling, can lift the degeneracy between the three directions of the triplet order parameter. We neglect these effects in our calculations.

\section{PDW Instability: linearized analysis}\label{SecNoSOC}

The interactions in Eq. \ref{Eq:Hint} can lead to three types of particle-particle instabilities: a uniform superconducting instability (SC), and two pair density wave (PDW) instabilities, whose Cooper pairs have a total momentum of \(\mp 2 K=\pm K\), which we refer to as PDW\(_{\pm K}\) respectively. Uniform superconductivity has been analyzed in some detail in \cite{Shaffer20}, so here we focus on the PDW orders. 

\begin{figure}[htp]
	\centering
	\includegraphics[width=0.45\textwidth]{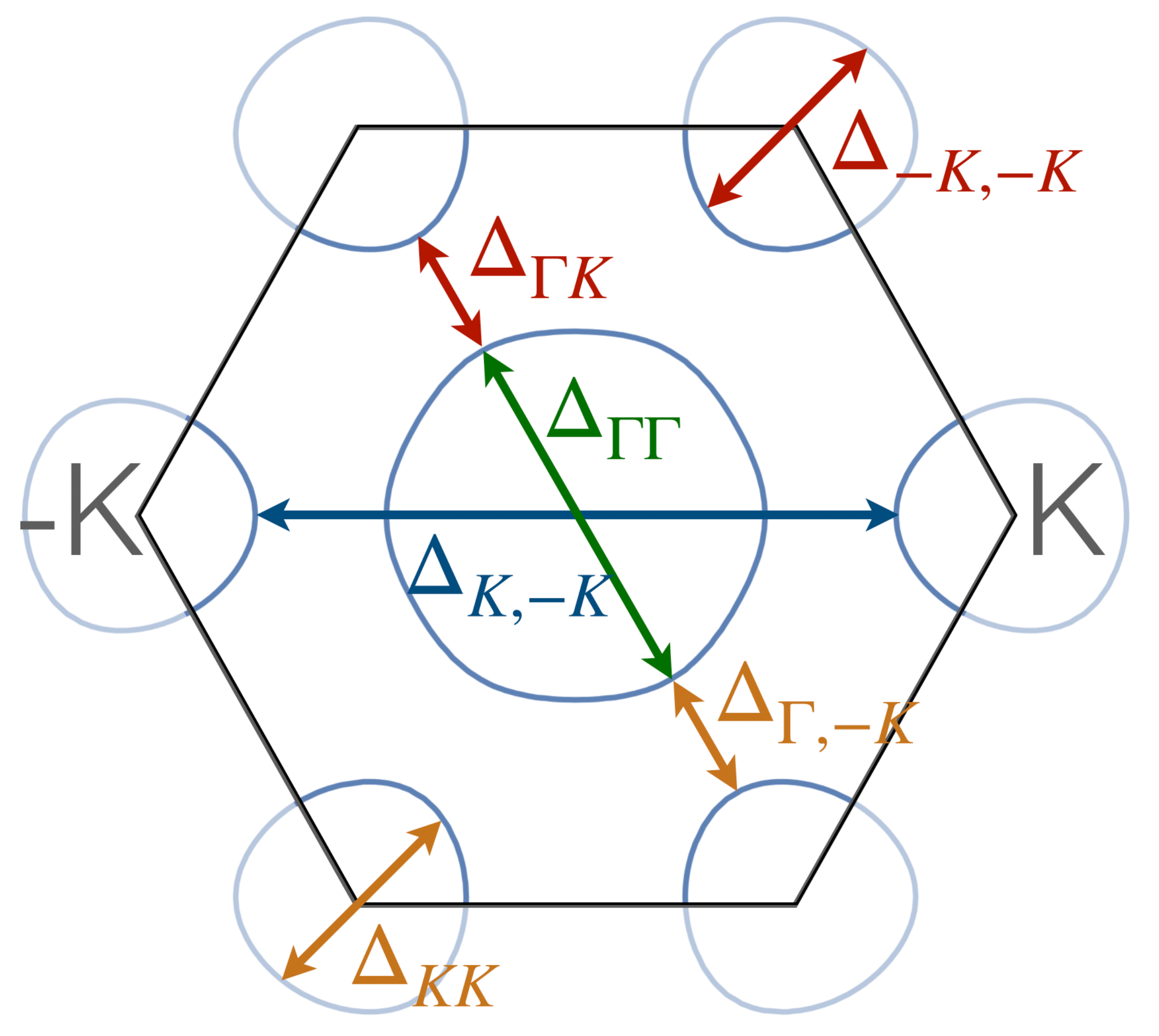}
	\caption{ \label{SCPDWorderFig}SC and PDW order parameters shown in the Brillouin zone. Green and blue are uniform SC, red is PDW\(_{-K}\), orange is PDW\(_{K}\).}
\end{figure}

The ordered phases are described by a pairing Hamiltonian (obtained from \(H_{\text{Int}}\) in Eq. \ref{Eq:Hint} by a Hubbard-Stratonovich transformation)
\[H_\Delta=\sum_{\mathbf{p}\eta\zeta\alpha\beta}\left[\widehat{\Delta}_{\eta\zeta}(\mathbf{p})\right]_{\alpha\beta}d^\dagger_{\mathbf{p}\eta\alpha}d^\dagger_{-\mathbf{p}\zeta\beta}+h.c.\label{HbdgPDW}\]
where the order parameters
\[\left[\widehat{\Delta}_{\eta\zeta}(\mathbf{p})\right]_{\alpha\beta}\propto \langle d_{\mathbf{p}\eta\alpha}d_{-\mathbf{p}\zeta\beta}\rangle\label{Delta}\]
are gap functions corresponding to uniform SC when \(\zeta + \eta = 0\), and to PDW\(_{\pm K}\) for \(\eta+\zeta=\pm K\).  In other words, a non-vanishing \(\widehat{\Delta}_{\eta\zeta}\)  indicates the formation of a condensate of pairs with total momentum \(\eta+\zeta\). In particular \(\widehat{\Delta}_{\Gamma\Gamma}\), \(\widehat{\Delta}_{K,-K}\) and  \(\widehat{\Delta}_{-K,K}\) are uniform SC gaps, while \(\widehat{\Delta}_{\pm K, \pm K}\),  \(\widehat{\Delta}_{\Gamma,\mp K}\) and \(\widehat{\Delta}_{\mp K,\Gamma}\) are PDW gaps with pair momentum \(\pm 2K=\mp K\) respectively, as summarized in Table \ref{SCPDWorders} and Fig. \ref{SCPDWorderFig}.  For convenience, we will also use $\widehat{\Delta}_{ K }$ ($\widehat{\Delta}_{- K }$) to denote any linear combination of PDW$_{K}$ (PDW$_{-K}$) gap functions, and $\widehat{\Delta}_0$ to denote any combination of uniform superconducting gaps.    

\begin{table}[htp] 
\begin{center}

\begin{tabular}{|c || c |c|}
\hline
SC  ( $\Delta_0$) & \(\Delta_{\Gamma\Gamma}\) & \(\Delta_{K,-K}, \Delta_{-KK}\) \\
\hline
PDW\(_{K}\)  ($\Delta_K$) & \(\Delta_{-K-K}\) & \(\Delta_{\Gamma K}, \Delta_{K\Gamma}\)\\
\hline
PDW\(_{-K}\)  ( $\Delta_{-K}$)& \(\Delta_{KK}\) & \(\Delta_{\Gamma,-K}, \Delta_{-K\Gamma}\) \\
\hline
\end{tabular}
\caption{SC and PDW order parameters.  In the right two columns, the two subscripts indicate which pockets are involved in the pairing, according to the definition in Eq. (\ref{HbdgPDW}).  In the first column, the subscript denotes the total momentum of the Cooper pair.  Thus $\Delta_{-K}$ denotes any linear combination of \(\Delta_{\Gamma K}, \Delta_{K,K},\), and \( \Delta_{K\Gamma}\)\, and similarly for $\Delta_{-K}$ and $\Delta_0$.}
\label{SCPDWorders}
\end{center} 
\end{table}

As we will see below, both spin-singlet \(\widehat{\Delta}_{\eta\zeta}(\mathbf{p})\propto i\sigma^y\) and spin-triplet \(\widehat{\Delta}_{\eta\zeta}(\mathbf{p})\propto \sigma^j i\sigma^y\) (\(j=x,y,z\)) orders are possible instabilities. The gaps satisfy particle-hole symmetry (PHS) : 
\[\widehat{\Delta}_{\eta\zeta}(\mathbf{p})=-\widehat{\Delta}_{\zeta\eta}^T(-\mathbf{p})\]
In the spin-singlet case this implies \(\widehat{\Delta}_{\eta\zeta}=\widehat{\Delta}_{\zeta\eta}\), while in the spin-triplet case, \(\widehat{\Delta}_{\eta\zeta}=-\widehat{\Delta}_{\zeta\eta}\). In particular, for momentum-independent gap functions, the spin-triplet gap necessarily involves pairing between different Fermi pockets (i.e. \(\eta\neq\zeta\)).

\subsection{Symmetries of the PDW states}
\label{PDWsym}

The pairing gaps identified above lead to a number of possible superconducting phases, which can be distinguished based on which symmetries are spontaneously broken.  While uniform SC pairing preserves translational symmetry, PDW breaks it: in real space on a triangular lattice, taking the inverse Fourier transform of Eq. (\ref{HbdgPDW}) shows that
the real space order parameters are proportional to \(e^{i\mathbf{Q}\cdot\mathbf{r}}\), where \(\mathbf{r}\) is the position in real space, and \(\mathbf{Q}\) is the total momentum of the Cooper pairs, which is 0 for SC and \(\pm \mathbf{K}\) for PDW\(_{\pm K}\) (see Appendix \ref{AppendixA}). From the symmetry point of view, the PDW\(_{\pm K}\) phases are therefore similar to the Fulde-Ferrell-Larkin-Ovchinikov (FFLO) phase  \cite{FF, LO} and both break a translational symmetry as well as the \(U(1)\) symmetry that is spontaneously broken by Cooper pair formation. Unlike FFLO, however, PDW emerges spontaneously in a material with no external fields breaking time reversal symmetry (TRS).

Consequently, the pairing Hamiltonian (\ref{HbdgPDW}) can have time-reversal symmetry (TRS), which acts as  \(T=i\sigma^y\mathcal{K}\), where \(\mathcal{K}\) is complex conjugation. Moreover, by combining TRS with  a global U(1) transformation  \(d^\dag \rightarrow e^{i\phi}d^\dag\), a whole family of anti-unitary symmetries \(T(\phi)=e^{i\phi}T\) can be generated; all of these square to \(-1\).  In the normal state, where the U(1) charge symmetry is  unbroken, $T(\phi)$ is a symmetry for every value of $\phi$.  In the uniform SC state, where U(1) is broken down to \(\mathbb{Z}_2\),  \(T(\phi)\) is a symmetry for only two choices of \(\phi\), whose specific values depend on the choice of phase of the order parameter.

For the PDW states of interest here, the situation is slightly more involved.  Under $T(\phi)$,  the pairing gaps transform as:
\[\widehat{\Delta}_{\pm K }(\mathbf{p})\xrightarrow[]{T(\phi)}e^{2i\phi} \sigma^y \widehat{\Delta}_{\mp K}^*(-\mathbf{p})\sigma^y\label{TRSeq}\]
A similar formula also applies to uniform SC gaps. TRS thus associates a PDW with pair momentum \(\eta+\zeta=\pm K\) with a PDW with opposite pair momentum \(\mp K\).    

When Cooper pairs with only one center-of-mass momentum \(\mathbf{Q}\) are present -- or more generally, whenever \(|\widehat{\Delta}_K|\neq |\widehat{\Delta}_{-K}|\) --  the resulting state spontaneously breaks TRS, since there is no value of $\phi$ for which Eq. (\ref{TRSeq}) leaves the pairing state invariant.  These PDW states also spontaneously break inversion symmetry, and expand the unit cell by a factor of three in real space (since \(3K=0\)), thereby reducing the translation symmetry as well.  Given the similarity with the state considered by Fulde and Ferrell \cite{FF}, we therefore follow Ref. \cite{AgterbergRev20} and denote this as a FF-type PDW phase.  Since inversion symmetry is broken in this state, mixing between spin-singlet and spin-triplet orders is in principle symmetry-allowed in this case \cite{AgterbergSigrist09}, though we  neglect this mixing below.

When Copper pairs with both momenta \(K \) and \(-K\) are present, i.e. \(|\widehat{\Delta}_K|=|\widehat{\Delta}_{-K}|\), time-reversal symmetry is preserved, since \(\phi\) can be taken to be half the relative phase between  \(\widehat{\Delta}_K\) and \(\widehat{\Delta}_{-K}\).  
The resulting pairing states also preserve inversion symmetry, and the real space order parameter is be proportional to \(\cos\left(  \mathbf{Q}\cdot\mathbf{r} +  \phi \right)\).  This form is similar to the state considered by Larkin and Ovchinikov \cite{LO}; we therefore follow the notation of Ref. \cite{AgterbergRev20} and refer to this as the LO-type PDW phase.   

At first sight, one might expect the existence of a second \(U(1)\) symmetry related to the relative phase between the two gap functions \(\widehat{\Delta}_K\) and \(\widehat{\Delta}_{-K}\). However, as we will show below, this \(U(1)\) symmetry  is broken down to \(\mathbb{Z}_3\) by non-linear terms in the gap equations.  
We can understand this residual  \(\mathbb{Z}_3\)  symmetry by examining translations within the extended unit cell. Under a translation \(\mathbf{T}_{\mathbf{a}_j}\) by a lattice basis vector \(\mathbf{a}_j\) with \(\mathbf{a}_1=(a,0)\) and  \(\mathbf{a}_2=\frac{a}{2}(1,\sqrt{3})\), noting that \(\mathbf{K}=\left(\frac{4\pi}{3a},0\right)\), the PDW order parameters transform as \cite{AgterbergRev20}
\begin{align}
\widehat{\Delta}_{\pm K}(\mathbf{p})&\xrightarrow[]{\mathbf{T}_{\mathbf{a}_1}}  e^{ \pm i \mathbf{a}_1\cdot \mathbf{K}}\widehat{\Delta}_{\pm K}(\mathbf{p})=e^{\mp \frac{2\pi i}{3}}\widehat{\Delta}_{\pm K}(\mathbf{p})\nonumber\label{TranslationEq}\\
\widehat{\Delta}_{\pm K }(\mathbf{p})&\xrightarrow[]{\mathbf{T}_{\mathbf{a}_2}}  e^{\pm i\mathbf{a}_2\cdot \mathbf{K}}\widehat{\Delta}_{\pm K }(\mathbf{p})=e^{\pm \frac{2\pi i}{3}}\widehat{\Delta}_{\pm K }(\mathbf{p})
\end{align}
Importantly, under translations the relative phase between \(\widehat{\Delta}_K\) and \(\widehat{\Delta}_{-K}\) shifts by \(\pm 4\pi/3=\mp 2\pi/3\), consistent with the fact that \(\mathbf{T}_{\mathbf{a}_j}\) are broken while  \(\mathbf{T}_{3\mathbf{a}_j}\) are not. 
Thus the residual \(\mathbb{Z}_3\) symmetry associated with the relative phase between \(\widehat{\Delta}_{-K}\) and \(\widehat{\Delta}_K\) simply reflects the fact that
 \(\mathbf{T}_{\mathbf{a}_j}\) is a symmetry of the Hamiltonian.  The resulting three-fold degeneracy ensures that states related to each other by translations by the original lattice basis vectors have the same energy. Evidently, we can choose \(\widehat{\Delta}_K=\widehat{\Delta}_{-K}\) in at most one of these three states; the other two are not invariant under the usual TRS $T$, but instead are symmetric under \(T(\phi)\) with \(\phi=\pm\pi/3\).  Since applying a lattice translation should not break TRS, it is natural to consider states symmetric under \(T(\phi)\) for any fixed \(\phi\) to be time-reversal invariant.

Finally, there are two additional symmetries of the gap functions that will be relevant to our analysis.  These are the spin interchange operation \(S\),  under which a spin-singlet (spin-triplet) gap function is odd (even),  and the momentum interchange operation \(P^*\), defined by:
\begin{align}
    \widehat{\Delta}_{\eta\zeta}(\mathbf{p})&\xrightarrow{S}\widehat{\Delta}^T_{\eta\zeta}(\mathbf{p})\nonumber\\
    \widehat{\Delta}_{\eta\zeta}(\mathbf{p})&\xrightarrow{P^*}\widehat{\Delta}_{\eta\zeta}(-\mathbf{p})
\end{align}
 Here  \(P^*\) is not to be confused with the parity \(P\) which additionally takes \(\eta,\zeta\rightarrow -\eta,-\zeta\).  
 To understand the significance of \(P^*\), we consider the time dependence of the gap function, \(\left[\widehat{\Delta}_{\eta\zeta}(\mathbf{p},t_1,t_2)\right]_{\alpha\beta}\propto \langle d_{\mathbf{p}\eta\alpha t_1}d_{-\mathbf{p}\zeta\beta t_2}\rangle\), and introduce the time interchange operation
\[\widehat{\Delta}_{\eta\zeta}(\mathbf{p},t_1,t_2)\xrightarrow{T^*}\widehat{\Delta}_{\eta\zeta}(\mathbf{p},t_2,t_1)\]
Note that \(T^*\) is distinct from TRS, which instead takes \(t_1,t_2\rightarrow -t_1,-t_2\).  Anti-commutation relations imply that \(SP^*T^*=-1\); hence the action of \(T^*\)  can be deduced from that of \(S\) and \(P^*\). Gap functions that are odd under \(T^*\) are the so-called odd-frequency gap functions \cite{GeilhufeBalatsky18, LinderBalatsky19}.  Although we only consider even-frequency gap functions as possible instabilities, we will see in Sec. \ref{InducedOrders} that odd-frequency uniform superconductivity can in principle be induced by even-frequency PDW.

In general, to resolve when the PDW realizes an FF- versus an LO-type order, as well as the other symmetry properties of the resulting phases, 
we will need to analyze the Ginzburg-Landau free energy beyond leading order, which we do in Sec. \ref{SecGL}. 

\subsection{Linearized Gap Equation}

\begin{table*}[htp] 
\begin{center}
\begin{tabular}{||c || c||}
\hline
\multicolumn{2}{c}{ Uniform SC channels}   \\
\hline
\begin{tabular}{c | cc}
\(g_{\eta,-\eta;\zeta,-\zeta}^{(s)}\) & \(\eta=\Gamma \) & \(\eta=\pm K \) \\
\hline
\(\zeta=\Gamma \) & \(g_1\) & \(g_4 \)\\
\(\zeta=\pm K\)& \(g_4\) & \((g_{2} + g_3)/2 \) 
\end{tabular}
 \ \ \ \ 
&
\begin{tabular}{c | cc}
\(g_{\eta,-\eta;\zeta,-\zeta}^{(t)}\) & \(\eta=\Gamma\) & \(\eta= \pm K\) \\
\hline
\(\zeta=\Gamma\) & \(g_1^{(t)}\) & \(g_4^{(t)}\)\\
\(\zeta=\pm K \)& \(g_4^{(t)}\) & \((g_{2} - g_{3})/2\) 
\end{tabular}
\ \ \ \ 
\\ 
\hline
 \multicolumn{2}{c}{ PDW channels} \\
 \hline
\begin{tabular}{c | cc}
\(g_{\eta,\eta' ;\zeta,\zeta'}^{(s)}\) & \(\eta=\Gamma, \eta'= \pm K \) &\ \  \(\eta=\eta'=\pm K\) \\
\hline
\(\zeta=\Gamma, \zeta'= \pm K \) &\((g_{6} + g_{7})/2\)   & \(g_8\)  \\
\(\zeta=\zeta'= \pm K \)&  \(g_8\) & \(g_5\)
\end{tabular}
\ \ \ \ 
&
\begin{tabular}{c | cc}
\(g_{\eta,\eta' ;\zeta,\zeta'}^{(t)}\) & \(\eta=\Gamma, \eta'= \pm K \) & \ \ \(\eta=\eta'=\pm K\) \\
\hline
\(\zeta=\Gamma, \zeta'= \pm K  \)  &  \((g_{6} - g_{7})/2 \)& \(g_8^{t}\)  \\
\(\zeta=\zeta'=\pm K \) &  \(g_8^{t}\) & \(g_5^{t}\)
\end{tabular}\\ 
\hline
\end{tabular}
\caption{Definition of the coupling constants \(g^{(\mu)}_{\eta\zeta;\eta'\zeta'}\) (\(\mu=s,t\) for singlet or triplet) in terms of the coupling constants \(g_n\) (\(n=1,\dots,8\)) introduced in Eq. (\ref{Eq:Hint}), and  $g^{t}_n$ ($n=1,4,5,8)$ introduced in Eq. (\ref{Eq:IntsKdep}).   
For the PDW channels, the couplings are invariant under $\eta \leftrightarrow \eta'$, $\zeta \leftrightarrow \zeta'$, so not all possible combinations are included explicitly.
}
\label{gTable}
\end{center} 
\end{table*}

To determine which PDW (if any) constitutes the system's dominant pairing instability,  we use a microscopic mean field theory. In this approach, the gap functions in Eq. \ref{Delta} satisfy the self-consistent gap equations:
\[\label{LinGapEqPDW}
\left[\widehat{\Delta}_{\eta\zeta}(\mathbf{p})\right]_{\alpha\beta}=-\sum_{\substack{\mathbf{k}\eta'\zeta'\\ \alpha'\beta'}}V^{\eta'\zeta';\alpha'\beta'}_{\eta\zeta;\alpha\beta}(\mathbf{p;k})\Pi_{\eta'\zeta'}(\mathbf{k}) \left[\widehat{\Delta}_{\eta'\zeta'}(\mathbf{k})\right]_{\alpha'\beta'}\]
where we include both the momentum-independent interactions (\ref{Eq:Ints_SpinChannel}) that drive the pairing instability, and small momentum-dependent interactions (\ref{Eq:IntsKdep}) needed to fully determine the gap function when the triplet instability dominates.

Here we consider the linearized gap equation, which is valid close to the phase transition.  In this case the (angle-resolved) pairing susceptibility, given by the particle-particle bubble, is independent of the gap functions:
\begin{align}\label{Eq:PiAng}
\Pi_{\eta\zeta}(\theta)&=-\frac{1}{ 8 \pi^2}\int
\frac{\tanh\left( \frac{\varepsilon_\eta(\mathbf{Q} )}{2T} \right) +\tanh\left ( \frac{\varepsilon_{\zeta}(-\mathbf{Q})}{2T} \right) }{\varepsilon_\eta(\mathbf{Q})+\varepsilon_{\zeta}(-\mathbf{Q})}QdQ\nonumber\\
&\approx\frac{N_{\eta}N_{\zeta}}{N_{\eta}+N_{\zeta}}\varpi\left(\epsilon_{\eta}(\mathbf{k})-\epsilon_{\zeta}(-\mathbf{k})\right)
\end{align}
where \(N_\eta=\frac{m_\eta}{4\pi}\) is the density of states (DOS) at the \(\eta\) pocket and we define the function
\[\varpi\left(x\right)=-\log\frac{1.13\Lambda}{T}+\text{Re}\left[\psi\left(\frac{1}{2}+\frac{i x}{4\pi T}\right)-\psi\left(\frac{1}{2}\right)\right]\label{varpiPDW}\]
with \(\psi\) being the digamma function and \(\Lambda\) the high energy cut-off. Note that $\epsilon_{\eta}(\mathbf{k})-\epsilon_{\zeta}(-\mathbf{k})$ depends only on the angle $\theta$, according to the dispersions in Eqs. (\ref{epsGamma}) and (\ref{epsK}).

Importantly, \(\varpi\left(0\right)=-\log\frac{1.13\Lambda}{T}\) implies a logarithmic instability, which means Eq. \ref{LinGapEqPDW} has a solution for arbitrarily weak interactions. We therefore refer to the condition \(\epsilon_{\eta}(\mathbf{k})= \epsilon_{\zeta}(-\mathbf{k})\), which results in a logarithmic instability, as perfect nesting. This condition is guaranteed  in the uniform SC channel (\(\eta=-\zeta\)), provided either time-reversal or inversion symmetries are present.   In the PDW channels, however, the trigonal warping \(w\) at the \(K\) pockets, Eq. (\ref{epsK}), and the Fermi surface mismatch between the \(\Gamma\) and \(K\) pockets both take the system away from the perfect nesting limit.  We therefore analyze the following two perturbations:
\[\epsilon_{K}(\mathbf{k})-\epsilon_{K}(-\mathbf{k})=2w\cos3\theta
\label{Eq:wdef} \]
(with \(\theta\) denoting the angle formed by \(\mathbf{k}\) with respect to the $\hat{x}$ axis) and 
\[\epsilon_{\Gamma}(\mathbf{k})-\epsilon_{K}(-\mathbf{k})=\hat{\mu}+w\cos3\theta\]
where we have linearized around the Fermi level and defined
\[\hat{\mu}=\frac{|m_\Gamma\mu_\Gamma-m_K\mu_K|}{m_\Gamma+m_K} \ .
\label{Eq:mudef} \]
Here, we also assume that the masses at the $\Gamma$ and $K$ pockets are comparable, \(\sqrt{m_\Gamma/m_K},\sqrt{m_K/m_\Gamma}\ll\Lambda/T_c\); otherwise additional corrections in the pairing susceptibility must be included.  We refer to \(w\) and \(\hat{\mu}\) as the detuning parameters, with \(w\) and \(\hat{\mu}\) both suppressing finite-momentum pairing between \(\Gamma\) and \(K\) pockets and \(w\) additionally suppressing finite-momentum pairing within the \(K\) pockets. Once either detuning parameter becomes non-zero, the pairing instability becomes a threshold instability, meaning that it only occurs for sufficiently strong interactions, a condition we will quantify below.

\subsubsection{Solutions of the Linearized Gap Equation}

To obtain the momentum-dependence of the solutions to the gap equation, we note that the interactions can be expressed in the form:
\begin{align}
[V^{(s)}]^{\eta'\zeta';\alpha'\beta'}_{\eta\zeta;\alpha\beta}(\theta,\theta')=&g^{(s)}_{\eta'\zeta';\eta\zeta} \left[\widehat{\Sigma}^{(s)}_{\eta\zeta}\right]_{\alpha\beta} \left[\widehat{\Sigma}^{(s)*}_{\eta'\zeta'}(\theta')\right]_{\alpha'\beta'} \nonumber \\
[V^{(t)}]^{\eta'\zeta';\alpha'\beta'}_{\eta\zeta;\alpha\beta}(\theta,\theta')=& g^{(t)}_{\eta'\zeta';\eta\zeta} \left[{\bf \widehat{\Sigma}}^{(t)}_{\eta\zeta}(\theta)\right]_{\alpha\beta} \cdot \left[ {\bf \widehat{\Sigma}}^{(t)*}_{\eta'\zeta'}(\theta')\right]_{\alpha'\beta'} \label{VintPDW} 
\end{align}
 Here  \(\mu =s\), \(t\)  denotes the spin-singlet (s) and spin-triplet (t) channels, and 
\begin{align}
\widehat{\Sigma}^{(s)}_{\eta\zeta}(\theta)&= i\sigma^y 
 \nonumber \\ 
{\bf \widehat{\Sigma}}^{(t)}_{\eta\zeta}(\theta)&=\boldsymbol{\sigma}  i\sigma^y  \Theta^{(t)}_{\eta\zeta}(\theta)  \ ,
\end{align}
where $\boldsymbol{\sigma}= (\sigma^x, \sigma^y, \sigma^z)$ is a vector of Pauli matrices. The functions \(\Theta^{(t)}_{\eta\zeta}(\theta)\) resulting from interactions that are uniform on all pockets are \(\pm1\) , while those arising from the momentum-dependent interactions in Eq. (\ref{Eq:IntsKdep}) are \(\sqrt2 \cos3\theta\), as summarized in Table \ref{TableSigma}.   The constants \(g^{(\mu)}_{\eta'\zeta';\eta\zeta}\), implicitly defined by Eqs. (\ref{Eq:Ints_SpinChannel}, \ref{Eq:IntsKdep}), are summarized in Table \ref{gTable}.

Plugging the expressions (\ref{VintPDW}) into Eq. (\ref{LinGapEqPDW}), we find that the momentum-dependence of the solutions to the linearized gap equation can be parametrized as:
\begin{align} \label{GapExpanPDW} 
\widehat{\Delta}^{(s)}_{\eta\zeta}(\mathbf{p}) =&\widehat{\Sigma}^{(s)}_{\eta\zeta}(\theta) \Delta^{(s)}_{\eta\zeta} \nonumber \\ 
\widehat{\Delta}^{(t)}_{\eta\zeta}(\mathbf{p}) =& \hat{{\bf d}} \cdot \boldsymbol{\widehat{\Sigma}}^{(t)}_{\eta\zeta}(\theta)\Delta^{(t)}_{\eta\zeta}
\end{align}
where \(\Delta^{(\mu)}_{\eta\zeta}\) are momentum-independent. The unit vector $\hat{{\bf d}}$ indicates the direction of the triplet pairing vector, which spontaneously breaks  spin-rotation symmetry; in the absence of SOC the free energy (and hence the phase diagram of interest here) does not depend on  $\hat{{\bf d}}$.

\begin{table}[htp] 
\begin{center}
\begin{tabular}{c | cc}
 & \(\quad\widehat{\Sigma}^{(\mu)}_{\eta\eta}\quad\) & \(\widehat{\Sigma}^{(\mu)}_{\eta\zeta\neq\eta}\) \\
\hline\hline
singlet & \(i\sigma^y\) & \(i\sigma^y\) \\
\hline
triplet & \(\sqrt{2}\cos3\theta \boldsymbol{\sigma} i\sigma^y\) & \(\epsilon_{\eta\zeta} \boldsymbol{\sigma} i\sigma^y\) \\
\end{tabular}
\caption{Basis functions \(\widehat{\Sigma}^{(\mu)}_{\eta\zeta}(\mathbf{p})\) for singlet (\(\mu=s\)) and triplet (\(\mu=t\)) channels in Eq. \ref{GapExpanPDW} for \(\eta=\zeta\) (intra-pocket) and \(\eta \neq \zeta\) (interpocket). $\boldsymbol{\sigma}$ is a vector of Pauli matrices, and \(\epsilon_{\eta\zeta}\) is the Levi-Civita symbol. \(\theta\) is the angle made by the small momentum \(\mathbf{p}\) measured with respect to the \(K\) direction.}
\label{TableSigma}
\end{center} 
\end{table}

To solve for the gaps $\Delta^{(\mu)}_{\eta \zeta}$, we exploit the fact that the equations for SC, PDW\(_{ K}\), and PDW\(_{ -K}\) channels, as well as those for singlet and triplet channels, are decoupled. This follows from momentum and spin conservation, respectively. Note that  since the FF-type PDW phase breaks inversion symmetry, in principle the singlet and triplet channels can mix \cite{AgterbergSigrist09}, but we expect one of the channels to be dominant and neglect this effect here.  
Inserting \ref{GapExpanPDW} back into the gap equation \ref{LinGapEqPDW} therefore yields six decoupled pairs of reduced gap equations, which can be solved to obtain singlet and triplet gaps in the
SC, PDW\(_{K}\), and PDW\(_{-K}\) channels.

The SC solutions have been analyzed in detail by us and Kang in \cite{Shaffer20},  and are reviewed in Appendix \ref{AppendixB} for convenience. In this paper, our focus is only on the PDW\(_{\pm K}\) channels, for which the reduced gap equations are
\[ \label{Eq:PDWGapSol}
\left(\begin{array}{c}
\Delta^{(\mu)}_{\pm K \pm K}\\
\Delta^{(\mu)}_{\Gamma, \mp K}
\end{array}\right)=\left(\begin{array}{ccc}
\hat{g}^{(\mu)}_5 & 2\hat{g}^{(\mu)}_{8K}\\
\hat{g}^{(\mu)}_{8\Gamma} & 2\hat{g}^{(\mu)}_{67}
\end{array}\right)\left(\begin{array}{c}
\Delta^{(\mu)}_{\pm K \pm K}\\
\Delta^{(\mu)}_{\Gamma, \mp K} 
\end{array}\right) \ .\]
We have absorbed the particle-particle susceptibility into the coupling constants for convenience, defining
\begin{align}\nonumber\label{hs2}
 \hat{g}_5^{(s)}&=\bar{\Pi}_{KK}^{(s)} g_5, \quad& \hat{g}_5^{(t)}&=\bar{\Pi}_{KK}^{(t)} g_5^{t}\nonumber\\
 \hat{g}_{67}^{(s)}&=\bar{\Pi}_{\Gamma, -K}^{(s)} (g_{6} + g_{7})/2,  \quad& \hat{g}_{67}^{(t)}&=\bar{\Pi}_{\Gamma, -K}^{(t)} (g_{6} - g_{7})/2 \nonumber\\
 \hat{g}_{8\Gamma}^{(s)}&=\bar{\Pi}_{KK}^{(s)}g_8, \quad& \hat{g}_{8\Gamma}^{(t)}&=\bar{\Pi}_{KK}^{(t)}g_8^{t}\nonumber\\
  \hat{g}_{8K}^{(s)}&=\bar{\Pi}_{-K\Gamma}^{(s)}g_8, \quad& \hat{g}_{8K}^{(t)}&=\bar{\Pi}_{-K\Gamma}^{(t)}g_8^{t}
\end{align}
Here,
\[\bar{\Pi}_{\eta\zeta}^{(\mu)}=\int \Pi_{\eta\zeta}(\theta)\text{Tr}\left[\widehat{\Sigma}^{(\mu)}_{\eta\zeta}(\theta)\widehat{\Sigma}^{(\mu)*}_{\eta\zeta}(\theta)\right]\frac{d\theta}{2\pi}\label{Pi}\]
are Fermi surface averages of \(\Pi_{\eta\eta'}(\theta)\) in Eq. \ref{Eq:PiAng}, weighted by traces of the basis functions $\widehat{\Sigma}^{(\mu)}_{\eta\zeta}(\theta)$ for the relevant symmetry channels. The solutions to the reduced gap equations are given by the eigenvectors of the  \(2\times2\) matrix in Eq. (\ref{Eq:PDWGapSol}), 
\[\left(\begin{array}{c}
\Delta^{(\mu\pm)}_{\pm K,\pm K}\\
\Delta^{(\mu\pm)}_{\Gamma,\mp K}
\end{array}\right)\propto\left(\begin{array}{c}
\kappa^{(\mu \pm)} - 2 \hat{g}^{(\mu)}_{67} \\
\hat{g}^{(\mu)}_{8\Gamma}
\end{array}\right) \ ,
\label{PDWevecs}\]
which are expressed in terms of the two eigenvalues:
\[ \label{Eq:Kappas}
\kappa^{(\mu\pm)}= \frac{1}{2} \left(\hat{g}^{(\mu)}_5+2\hat{g}^{(\mu)}_{67}\pm \sqrt{(\hat{g}^{(\mu)}_5-2\hat{g}^{(\mu)}_{67})^2+8\hat{g}^{(\mu)}_{8\Gamma}\hat{g}^{(\mu)}_{8K}} \right)\]

Near the phase transition, the dominant pairing instability corresponds to the solution that gives the highest \(T_c\). This, in turn, is determined by setting the corresponding eigenvalues to  \(\kappa^{(\mu\pm)}=1\). Because
\(\kappa^{(\mu+)}\) is always larger than \(\kappa^{(\mu-)}\), it always gives a higher \(T_c\). We therefore drop the minus solutions and drop the superscript in \(\kappa^{(\mu+)}\equiv\kappa^{(\mu)}\) below.  

Evidently, the two PDW\(_{\pm K}\) channels satisfy the same linearized gap equation, and thus have the same $T_c$. This degeneracy follows from the fact that the PDW\(_{\pm K}\) solutions are related by time reversal symmetry. As we will see in Sec. \ref{SecGL}, however, this degeneracy is lifted in the non-linearized gap equation, or equivalently by higher order (in powers of the gap function) terms in the free energy, which can result in the spontaneous breaking of TRS.

\subsubsection{PDW at Perfect Nesting}

To set the stage for a more general study of the PDW instabilities, we first discuss the case of perfect nesting, obtained by setting the detuning parameters to zero, \(w=\hat{\mu}=0\). 
 For simplicity, in the following we neglect the subleading momentum-dependent interactions in the triplet channel, and take the DOS on \(\Gamma\) and \(K\) pockets to be equal, \(N_{\pm K}=N_\Gamma=N\). In that case all the averaged particle-particle bubbles are equal, \(\hat{\Pi}_{\eta\zeta}=\Pi=-N\log\frac{1.13\Lambda}{T}\), and the largest eigenvalues of the linearized gap equation reduce to
\begin{align}
\kappa^{s}&=-\frac{\Pi}{2}\left(g_5+g_{6}+g_7-\sqrt{(g_5-g_{6}-g_{7})^2+8g_8^2 }\right)\nonumber\\
\kappa^{t}&=-\Pi \max\left[0,g_7-g_6\right]
\end{align}

The PDW will be the dominant instability only if the largest of the \(\kappa^{(\mu)}\) is greater than all the corresponding eigenvalues \(\gamma^{(\mu)}\) in the competing uniform SC channels. Note that the pair susceptibility $\Pi$ is the same for both uniform SC and PDW instabilities in the case of perfect nesting. Because the SC instability depends only on the coupling constants \(g_1, \dots, g_4\) (see Appendix \ref{AppendixB} for full expressions), a leading PDW instability at perfect nesting is possible for some values of \(g_5, \dots, g_8\), which in turn are determined by the details of the microscopic interactions.  
It is worth emphasizing, however, that unlike the uniform SC channel, in which electron-phonon interactions are known to give rise to effective on-site attractive interactions, we are not aware of an analogous mechanism in the PDW channels, and thus we expect that in the case of attractive interactions, uniform SC is generically the dominant instability.   For this reason, we discuss only  those PDW instabilities that occur in the presence of purely repulsive interactions.

To elucidate when an instability in the PDW channel exists, we first note that in order for a solution with  \(T_c>0\) to exist, \(\kappa^{\mu}\) must be positive. Since \(\Pi<0\) diverges logarithmically as $T \rightarrow 0$, the PDW is a weak-coupling instability for perfect nesting. Therefore, because we assume repulsive interactions (for which \(g_j>0\)), an instability in the spin-singlet channel requires \(g_{8}^2>g_5(g_{6}+g_{7})\). In other words, \(g_8\) must be sufficiently large in magnitude compared to \(g_5\) and \(g_{6}+g_{7}\). In this case, Eq. \ref{PDWevecs} implies that \(\Delta^{(s)}_{KK}\) and \(\Delta^{(s)}_{\Gamma,-K}\) have opposite signs, resulting in a sign-changing ``$s$-wave" PDW state. This is an example of a case in which an effective attraction emerges from repulsive interactions due to the gap function changing sign, similar to the uniform \(s^{\pm}\) SC state considered in e.g. \cite{TrevisanRafael18, Mazin11}. In the spin-triplet channel, we have \(\kappa^{(t)}>0\) if and only if \(g_7>g_6\). In this case, the effective attraction is a result of the sign changing between \(\widehat{\Delta}_{\Gamma K}\) and \(\widehat{\Delta}_{K \Gamma}\), as they acquire opposite signs in the spin-triplet channel, giving rise to an ``$f$-wave" PDW state. 

The main conclusion is that, for repulsive interactions, non-trivial solutions of the linearized PDW gap equation exist only in the presence of a \(\Gamma\) pocket. This is because the PDW instabilities are promoted by either a large Umklapp interaction \(g_8\) (singlet PDW) or a large exchange interaction between \(\Gamma\) and \(K\) fermions, \(g_7\) (triplet PDW). Clearly, both of these processes involve fermions at \(\Gamma\). In the absence of the \(\Gamma\) pocket, which is the case for instance of doped 1H-MoS\(_2\), there is no effective attraction to first order in the interactions $g_j$. Nevertheless, as pointed out in \cite{KimNatCom17}, higher-order processes can lead to an effective attractive interaction in the triplet PDW channel, e.g. via the Kohn-Luttinger mechanism.

\begin{figure*}[htp]
	\centering
	\includegraphics[width=0.8\textwidth]{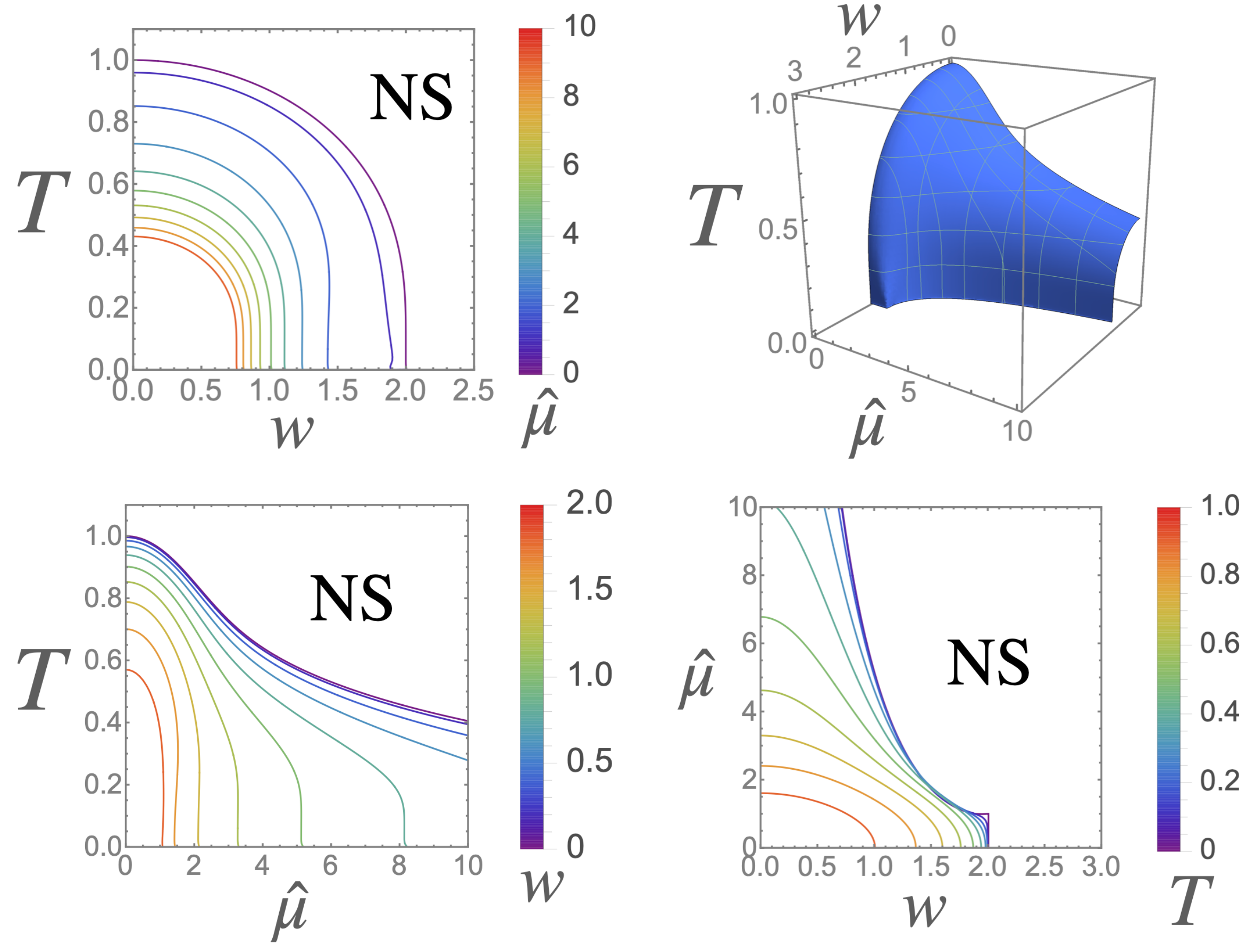}
	\caption{ \label{Fig:TcvsdetuningS}(color online). Critical surface for an instability of the singlet \((\mu=0)\) PDW channels (PDW\(_{\pm K}\) are degenerate) determined by Eq. (\ref{Eq:PDWGapSol}) in the phase space of temperature \(T\), trigonal warping \(w\), and Fermi surface mismatch between \(\Gamma\) and \(K\) pockets \(\hat{\mu}\), top right. Cuts are shown at constant \(\hat{\mu}\) (top left), constant \(w\) (bottom left), and constant \(T\) (bottom right) in units of \(T_{c0}\) (\(T_c\) at zero detuning). \(\mathrm{NS}\) indicates the normal state region of the phase diagrams. The coupling constants were set to \(g_6=1.2g_5\), \(g_8=2g_5\), \(g_7=1.05g_5\), \(g_5^t=0.2g_5\), and \(g_8^t=0.1g_5\).}
\end{figure*}

\begin{figure*}[htp]
	\centering
	\includegraphics[width=0.8\textwidth]{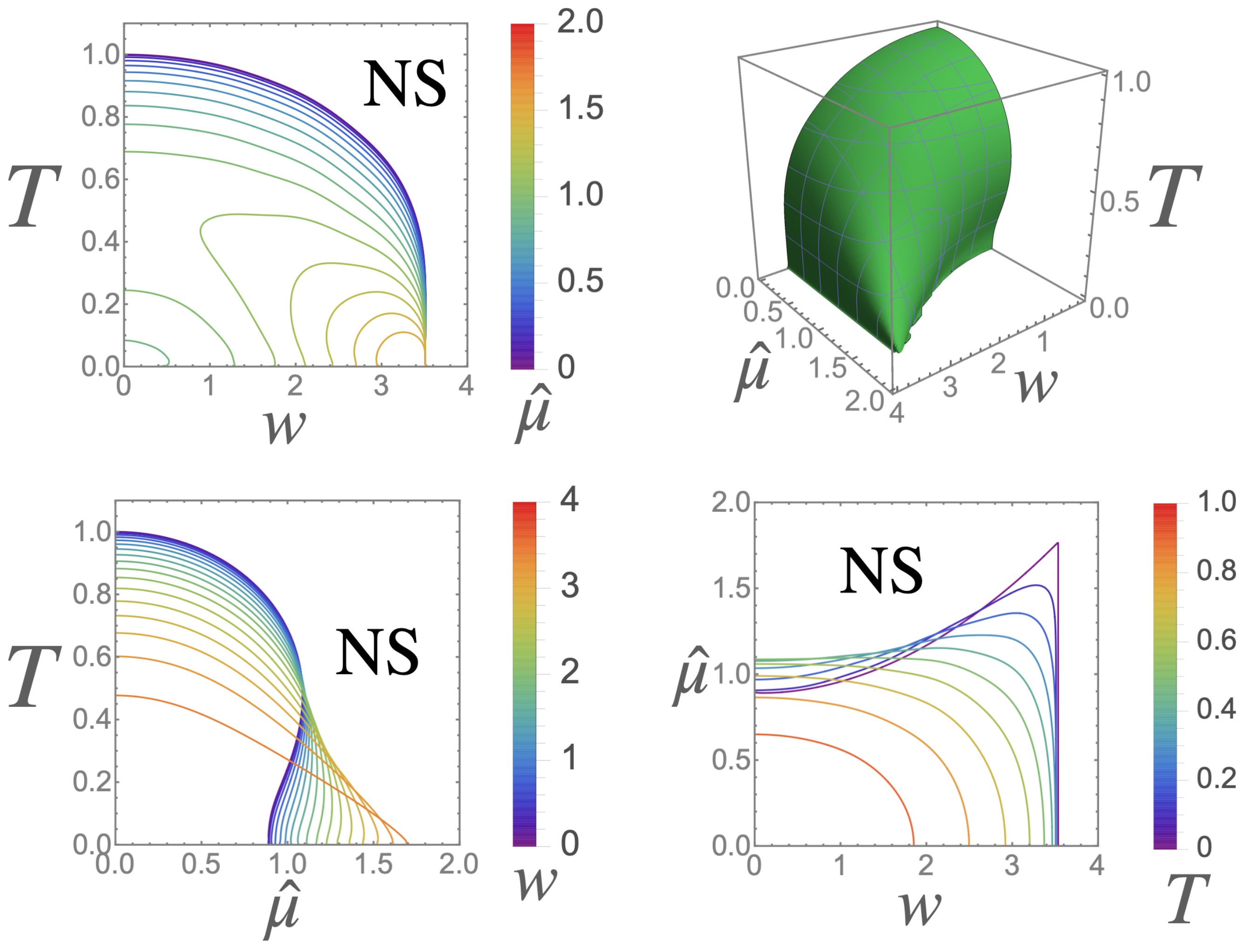}
	\caption{ \label{Fig:TcvsdetuningT}(color online). Same as Fig. \ref{Fig:TcvsdetuningS}, but for the triplet PDW instabilities \((\mu=x,y,z)\), which are degenerate in our model. All parameters are in units of \(T_{c0}\). The reentrant behavior in the \(T_c\) vs \(w\) plot (top left) is in part due to the non-single-valuedness of \(T_c\) as a function of \(\hat{\mu}\) that can be seen in the  \(T_c\) vs \(\hat{\mu}\) plot (bottom left), which indicates that a first order rather than a second order phase transition likely takes place for sufficiently large \(\hat{\mu}\). See also discussion in the text. \(\mathrm{NS}\) indicates the normal state region of the phase diagrams.  The coupling constants are the same as in Fig. \ref{Fig:TcvsdetuningS}, except for \(g_7=4.2g_5\).}
\end{figure*}

\subsection{Stability of the PDW away from perfect nesting}

Although PDW can be the instability for perfect nesting, in actual materials the Fermi surfaces are expected to be detuned from this limit.  We now analyze the stability of the PDW phase in the presence of the symmetry-allowed detuning  parameters \(w\) (trigonal warping of the \(K\) Fermi pocket) and \(\hat{\mu}\) (mismatch between \(\Gamma\) and \(K\) Fermi pockets) defined in Eqs. (\ref{Eq:wdef} - \ref{Eq:mudef}).  Both parameters enter into the susceptibility  $\bar{\Pi}_{\eta\zeta}$ (see Eqs. (\ref{Eq:PiAng}) and (\ref{Pi})), and generically reduce the eigenvalues \(\kappa^{(\mu)}\) in Eq. (\ref{Eq:Kappas}). Formally, these detuning parameters appear in the PDW pairing susceptibility, Eq. (\ref{varpiPDW}), in the same way as a Zeeman field appear in the uniform SC pairing susceptibility \cite{Maki64}. Therefore, the pair-breaking effect of the detuning of the Fermi pockets on the PDW transition temperature is equivalent to the pair-breaking effect of a magnetic field on the uniform SC transition temperature. Thus increasing the detuning lowers the PDW critical temperature \(T_c\).  At fixed temperature, the result is a critical surface in the space of detuning parameters, beyond which all $\kappa^{(\mu)} < 0$, and the PDW solution becomes unstable. 

The resulting phase diagrams in the parameter space of \(T, w,\) and \(\hat{\mu}\) are shown in the upper right panels of Figs. \ref{Fig:TcvsdetuningS} and \ref{Fig:TcvsdetuningT} for the PDW spin-singlet and spin-triplet phases, respectively.  \(T_c\) as a function of \(w\) (\(\hat{\mu}\)) for various fixed \(\hat{\mu}\) (\(w\)) is shown in the top (bottom) left panels. The critical values of the detuning parameters at which the PDW solution becomes unstable  at various fixed temperatures are shown in the bottom right panels; the values at \(T=0\) are obtained analytically. Since this is a multiband problem, the curves are not universal and depend (weakly) on the cutoff \(\Lambda\), which we set to \(2500 T_{c0}\), with \(T_{c0}\) being the critical temperature at zero detuning (\(w=\hat{\mu}=0\)).

These plots demonstrate that the PDW instability is robust for sufficiently small deformations of the Fermi surface away from perfect nesting.  Moreover, the singlet PDW is very robust to chemical potential mismatch between $\Gamma$ and $K$, while the triplet PDW is more robust to trigonal warping.

To understand the observed behavior of the PDW instability at finite detuning, it is enlightening to study the critical curves at zero temperature, which can be obtained analytically. In particular, the expressions for the weighted-average pairing susceptibilities at $T=0$ are
\begin{align}
\bar{\Pi}_{KK}^{(s)}&=-N_K\log\frac{\Lambda}{|w|}\nonumber\\
\bar{\Pi}_{KK}^{(t)}&=-N_K\log\frac{\Lambda}{e^{1/2}|w|}\nonumber\\
\bar{\Pi}_{\Gamma K}^{(\mu)}&=-\frac{2N_{\Gamma}N_K}{N_{\Gamma}+N_K} \log\frac{2\Lambda}{\left|\hat{\mu}+\sqrt{\hat{\mu}^2-\frac{w^2}{4}}\right|}
\label{Eq:Pi0det}
\end{align}
where \(\mu= s, t \) in the last line. The key implication of Eq. (\ref{Eq:Pi0det})  is that the detuning parameters effectively replace the temperature inside the logarithm, which implies in particular that there is a critical value of the trigonal warping parameter \(w_c\sim T_{c0}\)  above which the PDW phase become unstable at zero temperature, as can be seen from the purple curves in the bottom right panels of Fig. \ref{Fig:TcvsdetuningS} and \ref{Fig:TcvsdetuningT}. Note that \(T_{c0}\), which is the critical temperature for perfect nesting, has the same order of magnitude as the zero-temperature gap at perfect nesting, and it is set only by the strength of the interactions (here we work in units with the Boltzmann constant set to $1$). It follows that for fixed detuning \(w\), the interactions must be sufficiently strong for the PDW phase to be realized.  This is in stark contrast to the case of the logarithmic instability at perfect nesting, which can lead to pairing at zero temperature in the presence of arbitrarily weak interactions.

The parameter  \(\hat{\mu}\), on the other hand, only enters \(\bar{\Pi}_{\Gamma,-K}^{(\mu)}\) and does not affect \(\bar{\Pi}_{KK}^{(\mu)}\). As a result, the spin-singlet PDW phase is very stable against increasing \(\hat{\mu}\) at small \(w\), and the critical value of \(\hat{\mu}\) at which \(T_{c}\) goes to zero is on the order of the cutoff \(\Lambda\) (as can be shown analytically). Physically, this is because condensation at the \(K\) pocket remains energetically favorable -- recall that the effective attraction within the \(K\) pocket in this case still arises from pair hopping, and all interactions here are repulsive.
More precisely, while the pair hopping terms $\hat{g}_{8 \Gamma}$ and $\hat{g}_{8 K}$ that drive the PDW instability in our model decrease (as does \(T_c\)), \(\hat{g}_5\) and \(\hat{g}_{67}\) decrease at the same rate, so that it is still the case that \(\hat{g}_{8 \Gamma} \hat{g}_{8 K}> \hat{g}_5\hat{g}_{67}\), i.e. the stability criterion is satisfied. From Eq. (\ref{PDWevecs}) one can see that \(\Delta_{\Gamma K}\) decreases relative to \(\Delta_{KK}\)  as \(\hat{\mu}\) increases, so that the pairing within the \(K\) pocket remains energetically favorable since \(\bar{\Pi}_{KK}^{(\mu)}\) is unchanged.

Additionally, \(\bar{\Pi}_{\Gamma,-K}^{(\mu)}\) at zero temperature is a constant function of \(\hat{\mu}\) as long as \(\hat{\mu}<w/2\); this results in straight lines in the corresponding plots in Figs. \ref{Fig:TcvsdetuningS} and \ref{Fig:TcvsdetuningT}. We emphasize that this is a particular feature of our model and not a universal behavior, but similar non-analyticities are generically present in \(\bar{\Pi}_{\Gamma,- K}^{(\mu)}\) for other interaction and detuning parameterizations.

\begin{figure*}[htp]
	\centering
	\includegraphics[width=0.9\textwidth]{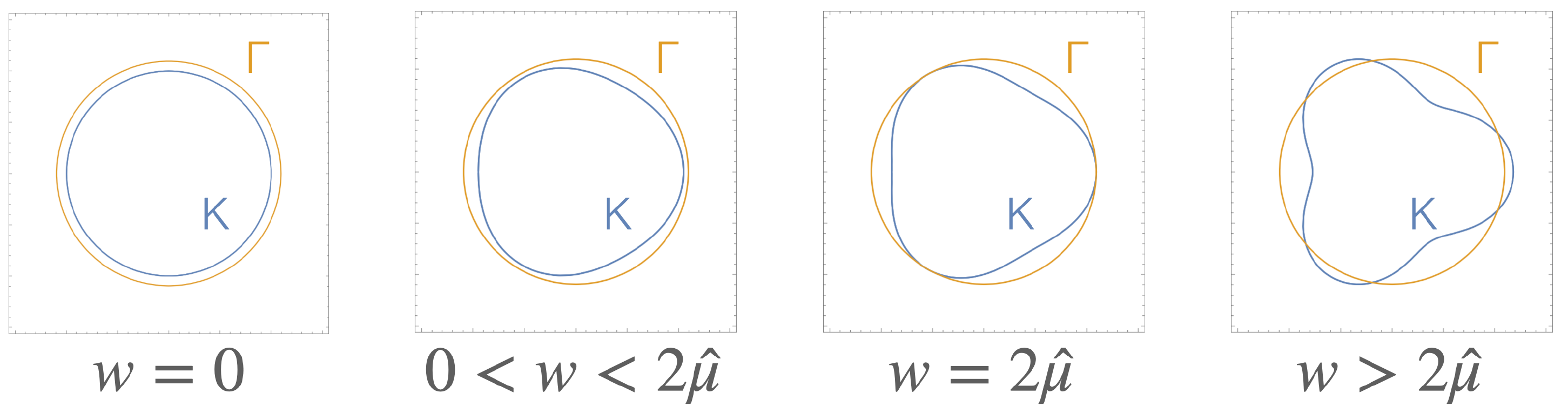}
	\caption{ \label{Fig:NestingFSvsTW} Schematic Fermi surfaces at the \(\Gamma\) (orange) and \(K\) (blue) points shifted to a common center for various values of trigonal warping \(w\) at fixed chemical potential mismatch, \(\hat{\mu}>0\) (color online). Nesting between the two Fermi surfaces is optimal when \(w=\hat{\mu}\).}
\end{figure*}

In the triplet case, in contrast, the only momentum-independent coupling in Eq. (\ref{hs2}) is $\hat{g}_{67}^t$; $\hat{g}_5^t$ and $\hat{g}_8^t$ correspond to subleading \(f\)-wave terms in the interactions, and are necessarily small.   \(\widehat{\Delta}_{KK}^{(t)}\) is therefore small, and the PDW is mostly stabilized by paring between \(\Gamma\) and \(K\) fermions, i.e. by \(\widehat{\Delta}^{(t)}_{\Gamma,-K}\). Consequently, the triplet channel is much more sensitive to the Fermi surface mismatch between the $\Gamma$ and $K$ pockets (parametrized by \(\hat{\mu}\)), such that \(T_{c}\) drops off sharply as \(\hat{\mu}\) increases.   In this case we find a critical value of \(\hat{\mu}\) at zero temperature, as shown in the bottom left panel of Fig. \ref{Fig:TcvsdetuningT}. In fact, \(T_c\), which was computed under the assumption of a second-order phase transition, is not even a single-valued function of \(\hat{\mu}\), as seen most clearly for the curve corresponding to \(w=0\) (purple). This suggests that the phase transition becomes first-order beyond the threshold value of \(\hat{\mu}\) where multiple solutions appear.  Mathematically, the problem is analogous not only to the standard FFLO instability of a uniform SC state with Zeeman-split bands \cite{FF, LO}, but also to an SDW instability of a system with detuned parabolic bands \cite{AndreySCSDWlikeFFLO}. In the latter case, the putative first-order phase transition turns out to be preempted by an SDW whose ordering wave-vector is incommensurate, i.e. distinct from the momentum separating the two pockets. An interesting possibility, which is beyond the scope of our work, is that a similar phenomenon may occur in our model, resulting in an incommensurate triplet PDW.

We also note that the critical value of \(\hat{\mu}\) necessary to completely suppress the triplet PDW state increases with increasing \(w\), suggesting that trigonal warping actually stabilizes the triplet PDW against the Fermi surface mismatch. As a result, there can be a re-entrant phase transition into the triplet PDW as either \(T\) or \(w\) increase. In the latter case, this happens because \(\bar{\Pi}_{\Gamma K}\) actually increases with \(w\) when \(\hat{\mu}>w/2\), which is when the Fermi surfaces at \(\Gamma\) and \(K\) (shifted to a common center) no longer overlap (see Fig. \ref{Fig:NestingFSvsTW}). As \(w\) increases, some points on the two shifted Fermi surfaces come closer, which increases \(\bar{\Pi}_{\Gamma K}\). The reentrance with increasing \(T\), on the other hand, is unphysical and indicates that a first order phase transition likely takes place at lower temperatures, as discussed above.

The main conclusion of this analysis is that the detuning of the Fermi surfaces -- trigonal warping of the $K$ pockets and mismatch between the $K$ and $\Gamma$ pockets -- act as pair-breaking for the PDW state, analogously to how a Zeeman magnetic field is pair-breaking for the uniform SC state. Importantly, these two detuning parameters, which are expected to be non-zero in a realistic system, impact the singlet and triplet PDW instabilities differently, with singlet PDW being much more robust to chemical potential mismatch between $\Gamma$ and $K$ pockets. However, as long as the energy scales associated with the detuning are small compared to the bare transition temperature $T_c$, the PDW instability can still be driven by weak to moderate interactions.

\section{PDW order: Ginzburg-Landau Free Energy Analysis}\label{SecGL}

The mean-field analysis of the linearized gap equations is insufficient to determine all the symmetry properties of the PDW phases.  This is because our analysis only shows that the PDW\(_{K}\) and PDW\(_{-K}\) orders are degenerate (as expected from symmetry considerations), without however setting the relative amplitudes and phases between them. This information is necessary to determine whether the PDW is of FF- or LO-type (and thus whether or not TRS is spontaneously broken) as well as to establish whether the apparent $U(1)$ symmetry associated with the relative phase rotation between the two PDW gaps remains unbroken.

To address these questions, rather than analyzing the non-linear gap equations, we study the Ginzburg-Landau free energy $\mathcal{F}$ to higher order in the order parameters.  Although the discussion that follows is predominantly phenomenological,  $\mathcal{F}$ can be derived from the microscopic theory.  We carry out this analysis in Appendix \ref{AppendixC}, and will utilize some of the results later to determine the signs and relative magnitudes of the  Ginzburg-Landau coefficients.  

To simplify our expression for $\mathcal{F}$, we express our free energy in terms of the gaps $\Delta_{\mp K}^{(\mu)}$ and $ \Delta_{0}^{(\mu)}$ associated with the total pairing gaps of the PDW$_{\pm K}$ and SC channels, respectively.  
Using this notation we can write the free energy as
\[\mathcal{F}=\sum_\mu \left[a_0^{(\mu)}|\Delta^{(\mu)}_0|^2+a_K^{(\mu)}\left(|\Delta_K^{(\mu)}|^2+|\Delta_{-K}^{(\mu)}|^2\right)\right]+\mathcal{F}^{(4)}+\dots\label{F}\]
where \(a_0^{(\mu)} \propto T - T_{\mathrm{SC}}^{(\mu)}\) and \(a_K^{(\mu)} \propto T_{\mathrm{PDW}}^{(\mu)} \) are coefficients that depend on the coupling constants, temperature, and detuning parameters.

To connect this expression back to the pairing gaps $\Delta_{\eta, \zeta}$ discussed in our mean-field analysis, we fix:
\begin{eqnarray}
\Delta_{\pm K,\pm K}^{(\mu)}&=r^{(\mu)} \Delta_{\Gamma,\mp K}^{(\mu)}\equiv \Delta_{\mp K}^{(\mu)}\nonumber\\
\Delta_{\Gamma\Gamma}^{(\mu)}&=r^{(\mu)}_0 \Delta_{K,-K}^{(\mu)}\equiv \Delta_{0}^{(\mu)}
\end{eqnarray}
where $r^{(\mu)}$ is the same for both PDW channels by symmetry. The ratios $r^{(\mu)}$ of the different contributions to $\Delta_{\pm K}$ are fixed by the relevant solution to the self-consistent gap equation. While in general \(r^{(\mu)}\) can be complex, close to the phase transition, it is given to leading order by Eq. \ref{PDWevecs}:
\[r^{(\mu)}=\frac{\kappa^{(\mu)} - 2 \hat{g}^{(\mu)}_{67} }{\hat{g}^{(\mu)}_{8\Gamma}}\label{rs} . \]

While higher order terms in the free energy can change this ratio, close to the phase transition they are subleading, and the difference is negligible. With this definition, the values of the coefficients \(a_0^{(\mu)}\) and \(a_K^{(\mu)}\) are fixed by requiring that minimizing the quadratic terms in $\mathcal{F}$ with respect to $\Delta_\eta^{(\mu)}$ returns the solutions of the reduced gap equations (\ref{Eq:PDWGapSol}).  Note that there is no mixing between singlet and triplet channels in the absence of SOC at all orders, so we assume only one of them is relevant close to \(T_c\).

Assuming unitary pairing, i.e. \(\mathbf{d}\times\mathbf{d}^*=0\), in the triplet channel the quartic part of the free energy depends only on the magnitude of the pairing vector, so that we need only use the scalar quantity $\Delta^{(t)}_\eta$ to parametrize the gap in this case (in general non-unitary pairing is allowed but not energetically favored, see Appendix \ref{AppendixC} for details). The quartic free energy for both singlet and triplet channels then has the form
\begin{widetext}
\begin{eqnarray}\label{F4PDW}
\mathcal{F}^{(4)}&=&\beta_1 |\Delta_0|^4 +\beta_2 \left(|\Delta_K|^4+|\Delta_{-K}|^4\right)+\beta_3 |\Delta_0|^2\left(|\Delta_K|^2+|\Delta_{-K}|^2\right)+\beta_4|\Delta_K|^2|\Delta_{-K}|^2+\nonumber\\
&+&\beta_5 \left(\Delta_0^2\Delta_K^*\Delta_{-K}^*+c.c.\right)+\beta_6 \left(\Delta_K^2\Delta_0^*\Delta_{-K}^*+\Delta_{-K}^2\Delta_0^*\Delta_{K}^*+c.c.\right)
\end{eqnarray}
\end{widetext}
where we include the uniform SC channel terms of the same parity and dropped the superscript $\mu$ for simplicity of notation. Note that the last term is allowed because $3K=0$.

From a phenomenological standpoint, we can treat \(\beta_n\) as variable parameters; in general, they depend on the details of the microscopic theory.  Depending on the values of these parameters, $\mathcal{F}^{(4)}$ can favor several different types of orders. For instance, the sign of $\beta_4$ determines whether both PDW gaps condense ($|\Delta_K| = |\Delta_{-K}|\neq0$), resulting in an LO-type phase, or whether only one of them condense, resulting in an FF-type phase. Quite generally, one expects $\beta_3 > 0$, indicating competition between uniform SC and PDW. Nevertheless, the Umklapp term with $\beta_6$ coefficient shows that the LO-type phase may induce uniform SC as a secondary order.

In the case where uniform SC and PDW orders coexist, the relative phases between the order parameters are determined by the signs and relative magnitudes of the parameters $\beta_5$ and $\beta_6$.  
Defining \(\Delta_\eta=|\Delta_\eta|e^{i\phi_\eta}\),   if $\beta_5$ is negative,  $\mathcal{F}^{(4)}$ has three degenerate minima.  These are given by \( \{ \phi_0=\phi_K=\phi_{-K}, \phi_0=\phi_K\pm\frac{2\pi}{3}=\phi_{-K}\mp\frac{2\pi}{3} \} \) for $\beta_6 <0$ and  \( \{ \phi_0=\phi_K+ \pi =\phi_{-K} + \pi, \phi_0=\phi_K\pm \frac{\pi}{3}=\phi_{-K} \mp \frac{\pi}{3} \} \) for $\beta_6 >0$.  Both the 3-fold degeneracy and the specific values are consistent with 3-fold translational symmetry breaking, which shifts the difference $\phi_K- \phi_{-K}$ by $\pm 2 \pi/3$ (see Sec. \ref{PDWsym}).  
 If $\beta_5 >0$, then no choice of the relative phases simultaneously renders both the terms with coefficients $\beta_5$ and $\beta_6$ negative, and the solutions depend on the ratio \(\beta_5/\beta_6\). However, the resulting ground states are always at least three-fold degenerate, consistent with the breaking of the three-fold translation symmetry.

While hereafter we analyze the behavior of the free energy for generic Ginzburg-Landau coefficients, as appropriate for any system with PDW instabilities with wave-vector $K$, for completeness we list the values of the \(\beta_n\) coefficients obtained from our model at perfect nesting (see Appendix \ref{AppendixC}):
\begin{align}\label{betas}
\beta_1/\beta_0&=2+cr_0^4\nonumber\\
\beta_2/\beta_0&=2+cr^4\nonumber\\
\beta_3/\beta_0&=4(1+r_0^2+r^2)\nonumber\\
\beta_4/\beta_0&=4(1+2r^2)\nonumber\\
\beta_5/\beta_0&=2(r^2+2(3-2c)r_0)\nonumber\\
\beta_6/\beta_0&=2r(r_0+2(3-2c))  \ .
\end{align}
Here, the ratios \(r=r^{(\mu)}\) are given in Eq. (\ref{rs}), $\beta_0=\frac{7\zeta(3)N}{32\pi^2T^2}$, with \(\zeta(3)\approx1.202\) the Riemann zeta function, and \(N\) the DOS, which we have taken to be equal on all pockets for simplicity. The coefficients $c$ are different in the singlet and triplet case: we find \(c=1\) in the singlet channel and \(c=3/2\) in the triplet channel.

\subsection{LO vs FF order in pure PDW phases}\label{SecLOvsFF}

We now consider the case where uniform superconductivity does not coexist with PDW, to analyze the possible symmetry-breaking patterns of the pure PDW phases.  
When $\Delta_0 = 0$, the free energy simplifies to:
\[\mathcal{F}^{(4)}_{PDW}=\beta_2 \left(|\Delta_K|^4+|\Delta_{-K}|^4\right)+\beta_4|\Delta_K|^2|\Delta_{-K}|^2\label{F4noSC}\]
To this order in the free energy, there are two possible ground states: if \(\beta_4>2\beta_2\), the ground state is an FF-type PDW (\(\Delta_{-K}=0\) or \(\Delta_{K}=0\)), which spontaneously breaks TRS; for \(\beta_4<2\beta_2\), the ground state is an LO-type PDW (\(|\Delta_{-K}|=|\Delta_{K}|\)), which preserves TRS.

Within our model, at perfect nesting, the sign of $\beta_4 - 2\beta_2$ depends only on the ratio $r$ between \(|\Delta_{KK}|\) and \(|\Delta_{\Gamma,-K}|\), which in turn is given by Eq. (\ref{rs}). Using Eq. \ref{betas}, we find a critical $r^* = 2$ ($r^* = \sqrt{6}$) such that, for $|r|<r^*$, the singlet (triplet) LO-type PDW is realized whereas for $|r|>r^*$, the singlet (triplet) FF-type PDW is realized. Thus, in the case of a triplet PDW, because $g_8^{(t)}$ is assumed to be small compared to \(g_{67}^{(t)}\)), $|r|$ is expected to be large, leading to an FF-type triplet PDW. Conversely, because the singlet PDW is favored when \(g_8\) is large, we expect it to be most likely of the LO-type. 

Although the particular critical values $r^*$ cited above are specific to our model, the  key qualitative observation is that if the dominant pairing is within the \(K\) pockets (small $|r|$), LO-type PDW is favored. This has a simple physical explanation: when pairing is predominantly intrapocket, i.e. within the \(K\) pocket, the dominant PDW\(_{K}\) and PDW\(_{-K}\) gaps are \(\widehat{\Delta}_{-K,-K}\) and \(\widehat{\Delta}_{KK}\) respectively with corresponding Cooper pairs formed from fermions from a single Fermi pocket.  As a result there is no competition between the two pairing channels (no trade off in the free energy to establish both orders), and the LO-type phase, where both orders are present with equal magnitude, is favored.  In contrast, if the dominant pairing is between \(\Gamma\) and \(\pm K\) pockets (meaning that the dominant PDW\(_{\pm K}\) gaps are \(\widehat{\Delta}_{\Gamma,\pm K}\)), the corresponding Cooper pairs in both channels involve the fermions at the \(\Gamma\) pocket, and the two channels compete.  As a result, it is energetically favorable for only one of the orders to be present at a time, and the FF-type phase is established instead.

The PDW free-energy up to quartic order does not depend on the relative phases between $\Delta_K$  and $\Delta_{-K}$. In fact, the relative phase of the two PDW gaps in the LO-type PDW is fixed only by a sixth-order terms in the free energy:
\begin{widetext}
\[\mathcal{F}^{(6)}_{PDW}=\Gamma_1 \left(|\Delta_K|^6+|\Delta_{-K}|^6\right)+\Gamma_2\left(|\Delta_K|^4|\Delta_{-K}|^2+|\Delta_{K}|^2|\Delta_{-K}|^4\right)+\Gamma_3\left(\Delta_K^3(\Delta_{-K}^*)^3+c.c.\right)\label{F6}\]
\end{widetext}
The \(\Gamma_3\) term is minimized when the relative phase satisfies \(\cos3(\phi_K-\phi_{-K})= 1\) for \(\Gamma_3<0\) and \(\cos3(\phi_K-\phi_{-K})= -1\) for \(\Gamma_3>0\). These two cases correspond, respectively, to \(\phi_K-\phi_{-K}=\frac{n\pi}{3}\) with even and odd integers \(n\). Consequently, the \(U(1)\) symmetry corresponding to the relative phase between the two PDW gaps is lowered to \(\mathbbm{Z}_3\). The three degenerate states map into each other under translation by a single lattice vector, reflecting the 3-fold translational symmetry breaking, as explained in Sec. \ref{PDWsym}. We summarize the results in Table \ref{PDWtable}.

\begin{table*}[htp]
\begin{center}
\label{orders}
\begin{tabular}{| c|c |c|c|c|c|c| c|}
\hline
  & \(S\) & \(P\)  & \(\mathbf{T}_{\mathbf{R}}\)  & \(U(1)\) & Induced Order (\(T^*\)) & PDW Type\\
\hline
\(\Delta_{\pm K}\) & \(-1\) & \(0\)  & \(0\) & \(2e\) & NA (1) & FF and LO\\

\(\Delta_{-K}\Delta^*_{K},\Delta_{K}\Delta^*_{0}\) & \(1\) & \(0\)  & \(0\)  & \(0e\) & \(\rho_K\) (1) & LO\\

\(\left|\Delta_{K}\right|^2-\left|\Delta_{-K}\right|^2\) & \(1\) & \(-1\)  & \(1\)  & \(0e\) & \(\ell\) (1) & FF\\

\(\Delta_{K}^2\Delta^*_{-K}+\Delta_{-K}^2\Delta^*_{K}\) & \(-1\) & \(1\)  & \(1\)  & \(2e\) & \(\Delta_0\) (1) & LO, \(\Gamma_3<0\) \\

\(\Delta_{K}^2\Delta^*_{-K}-\Delta_{-K}^2\Delta^*_{K}\) & \(-1\) & \(-1\) & \(1\)  & \(2e\) & \(\bar{\Delta}_0\) (-1) & LO, \(\Gamma_3>0\)\\

\(\Delta_{K}\Delta_{-K}\) & \(1\) & \(1\)  & \(1\) &  \(4e\) & \(\Delta_0^{(4e)}\) (1) & LO\\

\(\Delta_{K}^3+\Delta_{-K}^3\) & \(-1\) & \(1\)  & \(1\) & \(6e\) & \(\Delta_0^{(6e)}\) (1) & FF or LO, \(\Gamma_3<0\)\\

\(\Delta_{K}^3-\Delta_{-K}^3\) & \(-1\) & \(-1\)  & \(1\) &  \(6e\) & \(\Delta_0^{(6e)}\) (-1) & FF or LO, \(\Gamma_3>0\)\\

\hline
\end{tabular}

\caption{Symmetry properties of various products of \(\Delta_{\pm K}\) for the case of singlet PDW. The columns correspond to spin interchange \(S\), parity \(P\), translational symmetry \(\mathbf{T}_{\mathbf{R}}\), and the \(U(1)\) charge; the last two columns show the corresponding induced order, along with their sign under time-interchange \(T^*\) in parentheses,  and the PDW type that leads to the induces order, respectively. Entries of \(S,P,\mathbf{T}_{\mathbf{R}}=0\) imply the corresponding symmetry is broken by the combination. The possible induced orders are: \(\rho_K\) (charge density-wave); \(\ell\) (loop current); \(\Delta_0\) (uniform SC); \(\bar{\Delta}_0\) (odd-frequency) uniform SC); \(\Delta_0^{(4e)}\) (charge \(4e\) superconducting order); \(\Delta_0^{(6e)}\) (charge \(6e\) superconducting order); and \(\bar{\Delta}_0^{(6e)}\) (odd-frequency charge \(6e\) superconducting order). Whether even- or odd-frequency orders are induced in the LO-type PDW is determined by the sign of \(\Gamma_3\) in Eq. (\ref{F6}), as indicated in the last column. In the FF-type PDW, the induced 6e order is \(\Delta_0^{(6e)}\pm \bar{\Delta}_0^{(6e)}\).}
\label{PDWtable}
\end{center} 
\end{table*}

\begin{figure*}[htp]
	\centering
	\includegraphics[width=0.95\textwidth]{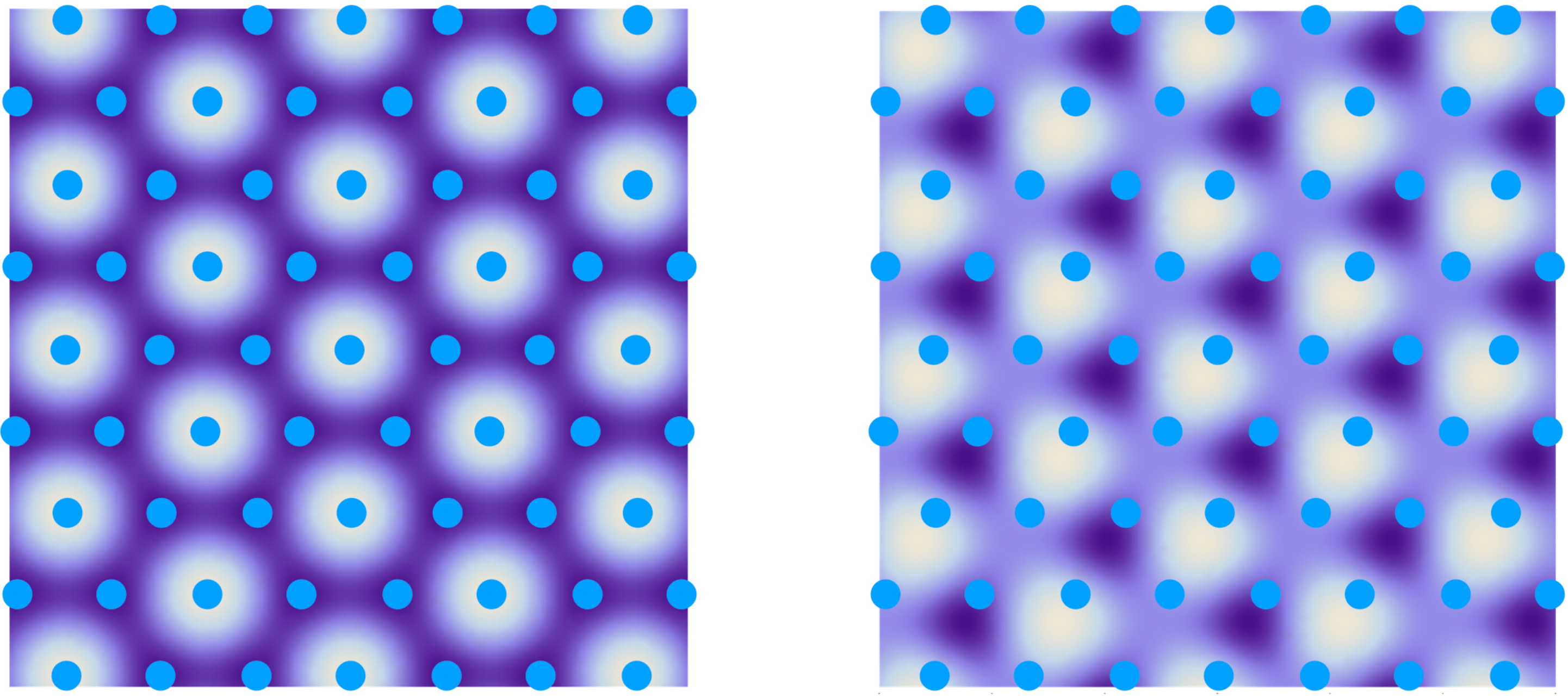}
	\caption{CDW induced by LO-type PDW in real space for \(\phi_K-\phi_{-K}=0\) and \(\pi/3\) respectively. Only the value at the lattice sites shown in blue are physical in the triangular lattice model.}\label{CDW}
	\label{CDWfig}
\end{figure*}

\subsection{Effect of uniform SC Fluctuations on the PDW}

Since the uniform SC and the PDW instabilities are driven by different interactions in our model, there is an interesting situation in which even though the PDW may be the leading instability, the uniform SC one is nearby. We now consider the fate of a system with a PDW instability that occurs at a temperature above, but close to, the uniform SC instability (\(a_0\gtrsim0\)).   In this, case SC fluctuations can be significant.  To study their impact on the PDW instability, we integrate out the SC fluctuations in the partition function:
\[\mathcal{Z}_{PDW}=\int e^{-\beta\mathcal{F}}\mathcal{D}[\Delta_0] \ . \]
Here $\mathcal{F}$ is  the free energy  to  quartic order in PDW fields \(\Delta_{\pm K}\) and to quadratic order in the SC field \(\Delta_0\), given in Eqs. \ref{F} and \ref{F4PDW} with \(\beta_1 = 0\). The \(\beta_6\) term, linear in \(\Delta_0\), gives rise to terms that are at least sixth-order in \(\Delta_{\pm K}\), and so we can drop this term as well. The resulting free energy can be re-written in terms of the real and imaginary parts of \(\Delta_0=\Delta'_0+i\Delta''_0\):
\begin{align}
\mathcal{F}=&a_K|\Delta_K|^2+a_K|\Delta_{-K}|^2+\beta_2\left(|\Delta_{K}|^4+|\Delta_{-K}|^4\right)+\nonumber\\
&+\beta_4|\Delta_{K}|^2|\Delta_{-K}|^2+\left(\begin{array}{c}
\Delta'_0\\
\Delta''_0
\end{array}\right)\cdot A\left(\begin{array}{c}
\Delta'_0\\
\Delta''_0
\end{array}\right)
\end{align}
where
\begin{align}
A=&(a_0+\beta_3\left(|\Delta_{K}|^2+|\Delta_{-K}|^2\right))\sigma^0+\nonumber\\
&+2\beta_5\text{Re}[\Delta_{K}\Delta_{-K}]\sigma^z+\beta_5\text{Im}[\Delta_{K}\Delta_{-K}]\sigma^x
\end{align}
is a real symmetric matrix with determinant
\[\det[A]=(a_0+\beta_3\left(|\Delta_{K}|^2+|\Delta_{-K}|^2\right))^2-4\beta_5^2|\Delta_{K}\Delta_{-K}|^2\]

We then perform the Gaussian integral to get
\[\mathcal{Z}_{PDW}=e^{-\beta \tilde{\mathcal{F}}},\]
using
\begin{align}
\int&\exp\left[-\beta\left(\begin{array}{c}
\Delta'_0\\
\Delta''_0
\end{array}\right)\cdot A\left(\begin{array}{c}
\Delta'_0\\
\Delta''_0
\end{array}\right)\right]d\Delta'_0 d\Delta''_0=\nonumber\\
&=\sqrt\frac{\pi^2}{\beta^2\det[A]}=\pi e^{-\frac{1}{2}\ln \beta^2 \det[A]}
\end{align}
Expanding to fourth order in \(\Delta_{\pm K}\), we have (dropping irrelevant constants)
\begin{align}
\tilde{\mathcal{F}}&=\left(a_K+\frac{\beta_3}{\beta a_0}\right)\left(|\Delta_K|^2+|\Delta_{-K}|^2\right)+\nonumber\\
&+\left(\beta_2-\frac{1}{2\beta}\left(\frac{\beta_3}{a_0}\right)^2\right)\left(|\Delta_{K}|^4+|\Delta_{-K}|^4\right)+\nonumber\\
&+\left(\beta_4-\frac{1}{\beta}\frac{\beta_3^2+2\beta_5^2}{a_0^2}\right)|\Delta_{K}|^2|\Delta_{-K}|^2
\end{align}
When the two transitions are in  proximity, (i.e. $a_0$ is small when $a_K=0$), SC fluctuations are in general detrimental to the PDW instability, since \(\beta_3\) is a positive definite term within our model. This is not unexpected since both orders are made up of the same fermions. 

Besides suppressing the PDW instability, SC fluctuations also mediate an effective coupling between PDW\(_{K}\) and PDW\(_{-K}\) orders.   Since this coupling is negative-definite, SC fluctuations tend to favor time-reversal invariant LO- type PDW phases over the time-reversal symmetry breaking FF ones.   This effect is particularly important in systems without a \(\Gamma\) pocket, for which the microscopic theory yields \(\beta_4=0\) (see Appendix \ref{AppendixC}), such that the two PDW order parameters would otherwise decouple in the free energy at quartic order, leaving the question of whether the PDW pairing is FF or LO unresolved at this order in the free energy.

\subsection{Induced Orders} \label{InducedOrders}

In general, the symmetry-breaking associated with both FF and LO -type PDW order parameters implies that in the corresponding ordered phases, additional induced orders will also exist \cite{AgterbergRev20,Agterberg08, BergFradkinKivelson09,WangAgterbergChubukovPRL15,WangAgterbergChubukovPRB15,BrydonAgterberg19}. Such induced orders are important as they can in principle be used to detect and distinguish the PDW phases in experiment if measured.
The possible induced orders corresponding to the FF- and LO-type PDW phases, along with their symmetries, are summarized in Table \ref{PDWtable}, and include charge-density wave, loop-current order, as well as various types of Cooper pairing with charge $2e$, $4e$, and $6e$.  Here, we describe how these possibilities arise from symmetry-allowed terms in the Ginzburg-Landau free-energy describing the PDW phases.

First, as was noted above, the \(\beta_6\) term in Eq. (\ref{F4PDW}) is linear in \(\Delta_0\), and is therefore the leading  term involving uniform SC when PDW is the dominant instability at the mean-field level. As a result,  uniform SC  with \(\Delta_0\propto \Delta_K^2\Delta_{-K}^*+\Delta_{-K}^2\Delta_{K}^*\) can be induced by the LO-type PDW, as can be seen by minimizing the free energy with respect to \(\Delta_0^*\).   Uniform SC order cannot be induced in FF-type PDW phases, as it requires both \(\Delta_K\) and \(\Delta_{-K}\) to be non-zero. 

Interestingly, an order parameter with opposite parity, \(\bar{\Delta}_0\propto \Delta_K^2\Delta_{-K}^*-\Delta_{-K}^2\Delta_{K}^*\), can also be induced. 
As discussed in Sec. \ref{PDWsym}, all gap functions are necessarily odd under \(SP^*T^*\), where recall that \(S\), \(P^*\) and \(T^*\) are spin-, momentum- and time-interchange operations respectively. In the uniform SC case, \(P^*\) is equivalent to parity \(P\), and so we conclude that while \(\Delta_0\) is even under \(T^*\), i.e. is the usual even-frequency order,  \(\bar{\Delta}_0\) must be an odd-frequency SC order \cite{Berezinskii74,VolkovEfetov03,TanakaSatoNagaosa12,BlackSchafferBalatsky13,TriolaBalatsky16,GeilhufeBalatsky18,LinderBalatsky19}.  It has been pointed out previously that such odd-frequency pairing can be induced by non-uniform paired states \cite{Higashitani14} such as PDW.

The question of whether the induced SC is even- or odd-frequency depends on the relative phase between \(\Delta_K\) and \(\Delta_{-K}\). Recall that \(\phi_K-\phi_{-K}=\frac{n\pi}{3}\) with even (odd) $n$, for negative (positive) $\Gamma_3>0$ in Eq. (\ref{F6}). Therefore, in the LO-type phase, we find:
\begin{align}
\Delta_0 & \propto |\Delta_K|^3 e^{i (\phi_K + n \pi/3)} (1+(-1)^n) \\
\bar{\Delta}_0 & \propto |\Delta_K|^3 e^{i (\phi_K + n \pi/3)} (1-(-1)^n)
\end{align}
i.e. $\Delta_0$ vanishes for positive $\Gamma_3$ (odd $n$) whereas $\bar{\Delta}_0$ vanishes for negative $\Gamma_3$ (even $n$). Thus, a singlet (triplet) PDW induces a singlet (triplet) SC gap that is even or odd frequency, depending on whether the sign of $\Gamma_3$ is negative or positive, respectively.  While we find that \(\Gamma_3\) is negative in the our simplified microscopic model, it could in principle be positive in more complex models, giving rise to a possibility that a pure odd-frequency uniform SC order can be induced by an LO-type PDW order.   
It is important to note, however, that if parity is broken (e.g. by Rashba or Ising spin-orbit coupling), which is the case in several monolayer TMDs, an even-frequency singlet (triplet) uniform SC gap necessarily accompanies the odd-frequency triplet (singlet) gap \cite{AgterbergSigrist09,TanakaSatoNagaosa12,Aliabad18,LinderBalatsky19}. The even-frequency contribution is likely to dominate the superconducting properties of the system.

\begin{figure}[htp]
	\centering
	\includegraphics[width=0.45\textwidth]{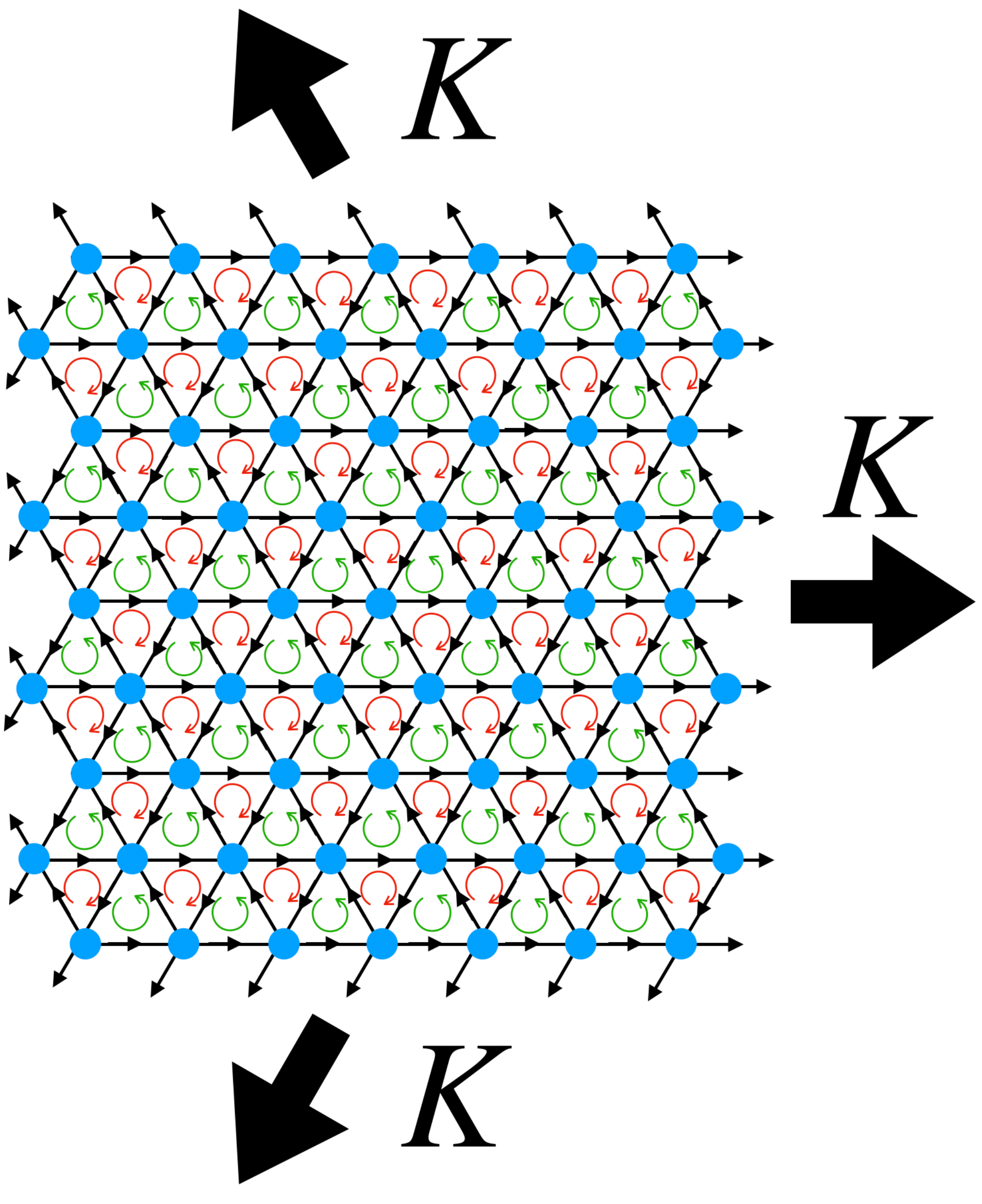}
	\caption{Schematic picture of PDW\(_{K}\) with total momentum \(K\) illustrated in real space where each arrow represents a jump in the complex phase of the order parameter by \(2\pi/3\) (see Appendix \ref{AppendixA}). One can schematically consider the arrows as induced currents along the three equivalent directions, giving rise to the loop currents shown in red and green. All arrows are reversed for PDW\(_{-K}\).\label{LoopFig}}
\end{figure}

Besides inducing uniform SC, the LO-type PDW phase also leads to a charge-density wave (CDW) order \(\rho_K\sim\langle d^\dagger_{\mathbf{p}\Gamma\alpha}d_{\mathbf{p},-K\alpha}\rangle,\langle d^\dagger_{\mathbf{p}K\alpha}d_{\mathbf{p},\Gamma,\alpha}\rangle\) via an additional term in the free energy \cite{Agterberg08, BergFradkinKivelson09,WangAgterbergChubukovPRL15,WangAgterbergChubukovPRB15, AgterbergRev20},
\[\mathcal{F}^{(3)}_{1}=\Lambda_{1}\rho_{K}\Delta_{K} \Delta^*_{-K}+c.c.\]
By definition, \(\rho_K=\rho_{-K}^*\), so \(\rho_{-K}\) is part of the same order parameter. Adding quadratic terms \(\mathcal{F}_\rho\propto \rho_{K}\rho^*_{-K}\) and minimizing with respect to \(\rho^*_{-K}\), we find that at the minimum of the free energy, a CDW order parameter \(\rho_{K}\propto \Delta_{-K} \Delta^*_{K}\) is induced in the LO-type PDW phase. Therefore, the phase of the CDW is the same as the relative phase of the two PDW order parameters, $\phi_K - \phi_{-K}$. Thus, the $Z_3$ symmetry of the relative phase corresponds to the different ways in which translational symmetry is broken in the LO-type PDW phase.

While the term above clearly vanishes for FF-type PDW orders, if uniform SC is present, mixing terms of the form
\[\mathcal{F}^{(3)}_{2}=\Lambda_{2}\rho_{K}\Delta_{-K} \Delta^*_{0}+c.c.\]
are also symmetry-allowed \cite{Agterberg15,Wang18,DaiSenthilLee18, Edkins19, AgterbergRev20}. For an FF-type PDW coexisting with a uniform SC phase, we thus obtain an induced CDW order \(\rho_{K}\propto \Delta_{K} \Delta^*_{0}\).

Interestingly, the FF-type PDW phase also induces loop-current order, described by the order parameter \(\ell\propto |\Delta_{K}|^2-|\Delta_{-K}|^2\). This type of induced order, absent in the LO-type phase,  \emph{preserves} translational symmetry (unlike the FF-type PDW itself) but still breaks TRS and inversion symmetry \cite{Lee02,AgterbergLoop15, AgterbergRev20}. Physically, it corresponds to a non-zero expectation value of the current operator along loops connecting nearest neighbor sites (see Fig. \ref{LoopFig}), which in this case is equivalent to a superposition of three currents along the three equivalent \(K\) directions. The broken inversion symmetry means that a type of effective spin-orbit coupling can also be induced in the FF-type PDW \cite{AgterbergSigrist09}, although in practice this is not the dominant source of SOC in 1H TMDs.

Finally, higher-order uniform superconductivity involving bound states of four or six fermions can be induced in LO-type PDW as \(\Delta_{4e}\propto \Delta_K\Delta_{-K}\) and \(\Delta_{6e}\propto \Delta_K^3+\Delta_{-K}^3\) \cite{Ropke98,Wu05, BergFradkinKivelson09,Agterberg11,Moon12,AgterbergRev20, FernandesFu21,GaraudBabaev21, MaccariBabaev22}.  The latter arises due to Umklapp processes, since \(3K=0\). An odd-frequency \(6e\) SC condensate of the form \(\bar{\Delta}_{6e}\propto \Delta_K^3-\Delta_{-K}^3\) can also be induced if \(\Gamma_3\) is positive. In FF-type PDW, only \(\Delta_{6e}\propto \Delta_K^3\) can arise. Note that bound states of \(4\) and \(6\) (or even more) fermions with non-zero momentum are also possible in principle \cite{Agterberg11}. We do not consider these higher order SC orders in detail.

\section{Discussion} \label{Sec:Discussion}

In this work, we showed that repulsive interactions can drive a weak-coupling PDW instability in systems with band structures similar to those of certain TMDs, like NbSe$_2$ and heavily doped MoS$_2$, where the Fermi surface consists of a pair of pockets at $\pm K$ and a pocket at $\Gamma$. In particular, among the 8 symmetry-allowed interactions involving the low-energy electronic states, 4 contribute only to the uniform SC instability whereas the other 4 contribute exclusively to the PDW instability. As a result, in a scenario in which repulsive interactions drive pairing and the pockets are perfectly nested, it is in principle possible for the PDW instability to win over the SC one. We also found that the PDW order remains stable even away from perfect nesting, as long as the energy scales of the detuning parameters (trigonal warping and mismatch between the Fermi momenta of the $\Gamma$ and $K$ pockets)  are not comparable with the pairing energy scale. Finally, we compared the secondary orders that are induced inside the PDW states, which involve not only charge order and loop-current order, but various types of uniform superconducting order with charges $2e$, $4e$, and $6e$.

Several of our results are worth highlighting. First, in our treatment, the $\Gamma$ pocket is in fact integral to the existence of a PDW instability, in both singlet and triplet channels.   Alternative mechanisms, such as the Kohn-Luttinger mechanism, could lead to such instabilities even in the absence of a $\Gamma$ pocket\cite{KimNatCom17}; however, they are not present at leading-order with generic spin-independent interactions.  Second, two types of PDW order can exist in these systems: FF-type, in which either the PDW with wave-vector $\mathbf{K}$ or the PDW with wave-vector $-\mathbf{K}$ condense, spontaneously breaking time-reversal symmetry; and  LO-type, which preserves time-reversal symmetry by condensing both PDW order parameters with wave-vectors $\mathbf{K}$ and $-\mathbf{K}$. While the former is favored when inter-pocket pairing between $K$ and $\Gamma$ dominates, the latter is favored when the dominant pairing channel is between electrons on the same $(\pm K)$ pocket.  Third, even when the PDW instability wins over the uniform SC instability, an LO-type PDW order is expected to induce uniform SC, such that both order parameters may simultaneously be present, resulting in a gap function that, while modulated, does not average to zero.

We now briefly discuss how our results are modified by the lack of inversion symmetry in 1H TMD monolayers. The model we considered in the main text has the same symmetries as the point group $D_{6h}$. However, in the case of a monolayer of 1H TMD, inversion symmetry is explicitly broken by the crystal structure, lowering the point group symmetry to $D_{3h}$. This results in the emergence of a so-called Ising SOC term in the Hamiltonian, whose impact on the uniform SC properties has been extensively discussed in the literature \cite{MakNat16,Exp3MoS2,Exp1MoS2,Exp2MoS2,Exp1NbSe2,Exp1TaS2,LawMay16,Houzet17,KimNatCom17,AjiPRB17,Oiwa18,KhodasMockliPRB20,Shaffer20,WickramaratneAgterbergMazin20,MargalitBerg21}.  Rather than pursue such a detailed analysis in the PDW case, here we focus on the qualitative effects of the Ising SOC.

For uniform SC, momentum reversal symmetry ensures that, in the presence of Ising SOC, pairing occurs predominantly between opposite spins, favoring singlet pairing and triplet pairing with a d-vector given by ${\bf d} = \hat{\mathbf{z}}$ \cite{FrigeriAgterberg04,LawPRB16,Khodas18,KhodasMockliPRB20,Shaffer20}. Importantly, because Ising SOC mixes this triplet state with the singlet state, the two comprise a single channel.  For PDW pairing, the impact of the SOC is clearer in the case of pairing within the \(\pm K\) pockets. This is because, for large SOC, intra-pocket pairing with momentum \(K\) can only occur between electrons of the same spin. This leads to the possibility of helical or chiral LO-type triplet PDW in the limit of strong SOC, as pointed out in \cite{KimNatCom17}. Note that in contrast to our work, Ref. \cite{KimNatCom17} considered a Kohn-Luttinger type mechanism for the attractive interactions between spin-polarized Fermi surfaces at \(K\) pockets, as that model, focusing on MoS$_2$, did not have a \(\Gamma\) pocket.
Within our microscopic model, however, the triplet PDW pairing occurs primarily between electrons on \(\Gamma\) and \(K\) pockets. Such a triplet PDW is expected to be of FF-type, suggesting rather a chiral triplet FF-type PDW if such an instability were to be realized. Further analysis, which we leave for future work, is required to properly account for the effects of SOC on the chirality of the gap functions when the \(\Gamma\) pocket is present, as well as in the case of spin-unpolarized \(K\) pockets.

Our simplified model, which is motivated by the band structure of certain TMDs, provides a possible route to realize unconventional PDW driven by repulsive electronic interactions. It remains to be seen whether actual TMD compound satisfies the conditions necessary for the PDW to be stabilized -- i.e. superconductivity driven by repulsive interactions, dominant $g_7$ and $g_8$ interactions, and Fermi pockets at $\pm K$ and $\Gamma$ of similar shapes and sizes. The recently reported PDW in NbSe$_2$ is unlikely to be related to the mechanisms discussed here, since the PDW wave-vector is along the $\Gamma$-$M$ direction \cite{LiuDavis20}. For the superconductors 2H-NbSe$_2$, monolayer NbSe$_2$, 2H-TaS$_2$, and doped monolayer MoS$_2$, the gap function is very likely conventional and predominantly s-wave like (although an admixture with f-wave takes place in the non-centrosymmetric monolayers due to Ising SOC \cite{KhodasMockliPRB20}), consistent with conventional electron-phonon pairing interaction \cite{Exp3MoS2,LawMay16,Dvir18}.
Yet, at least in few-layer NbSe$_2$, there is indirect evidence in favor of a subleading unconventional pairing channel, which would suggest the presence of electron-electron pairing interactions \cite{HamillPribiag20,Lortz20,WickramaratneAgterbergMazin20,KuzmanovicKhodas21,WanUgeda22}. The bulk compound 4Hb TaS$_2$, which has additional Fermi pockets at the $M$ point \cite{DentelskiFernandesRuhman21}, also displays superconducting properties typical of unconventional pairing mechanisms \cite{RibakRuhman20,PerskyRuhman22,AlmoalemRuhman22}. Besides the $s$-wave nature of the gap function, in the TMDs that contain a $\Gamma$ pocket, such as 2H-NbSe$_2$ and 2H-TaS$_2$, the pockets do not appear well nested \cite{Borisenko09,Zhao17}. Therefore, to potentially realize a PDW, it would be interesting to not only suppress the $T_c$ of the uniform SC state, but also to tune the band structure of these types of TMDs. While gating provides a powerful knob to tune the chemical potential of few-layer TMDs, another interesting possibility to manipulate the band structure of bulk TMDs is via the so-called misfit compounds $(MX)_{1+\delta}(TX_2)_n$ \cite{Rouxel95,Giang10,Trump14}. Here, the TMD layers, $TX_2$, are intercalated by monochalcogenide layers $MX$ with rock-salt structure. Charge is transferred between the rock-salt and TMD layers, which have different periodicity, thus changing the latter's band structure. Interestingly, several of these misfit compounds display superconductivity \cite{Oosawa92,RoeskyRouxel93,Nader97}. Given the large number of building blocks that can potentially be used in these misfit compounds, it would be interesting to search for combinations that can tune the band dispersion closer to the optimal conditions for a PDW to be observed.

\begin{acknowledgments}
We thank D. Agterberg and J. Kang for fruitful discussions. RMF was supported by the National Science Foundation through the University of Minnesota MRSEC under Award No. DMR-2011401. FJB was supported by the National Science Foundation under Award No. DMR-1928166. RMF thanks the hospitality of KITP during the program ``Recent Progress in the Experimental and Theoretical search for Pair-Density-Wave Order." KITP is supported in part by the National Science Foundation under Grant No. NSF PHY-1748958.
\end{acknowledgments}

\appendix

\section{Tight Binding Model}\label{AppendixA}

Here we present a simple tight-binging model that we used to produce the band structures in Fig. \ref{FigFS}, as well as study our order parameters in real space. The tight binding model is defined on the triangular lattice with sites we label with indices \(i\) and \(j\), which stand for \(\mathbf{R}_i=n_i \mathbf{a}_1+m_i \mathbf{a}_2\), where \(\mathbf{a}_1=(a,0)\) and  \(\mathbf{a}_2=\frac{a}{2}(1,\sqrt{3})\) are the lattice basis vectors. We take \(a=1\) for convenience below, and also define \(\mathbf{a}_3=\mathbf{a}_2-\mathbf{a}_1=\frac{a}{2}(-1,\sqrt{3})\). Each site then has six nearest neighbors at \(\pm\mathbf{a}_1\), \(\pm\mathbf{a}_2\) and \(\pm\mathbf{a}_3\). 

We describe our model in terms of the creation operators $d^\dag_{i, \alpha}$, where $\alpha =\uparrow, \downarrow$ is a spin index, and $i$ is a site index. The Hamiltonian has the form
\begin{align}
H_0&=\sum_{i\alpha}\mu\ d^\dagger_{i\alpha}d_{i\alpha}+\sum_{\langle ij\rangle\alpha}t\ d^\dagger_{i\alpha}d_{j\alpha}+\sum_{\langle\langle ij\rangle\rangle\alpha}t_{nnn}\ d^\dagger_{i\alpha}d_{j\alpha}
\end{align}
The next-nearest-neighbor hopping terms are needed to generate both \(\Gamma\) and \(\pm K\) pockets; the next-nearest-neighbors are located at \(\mathbf{a}_1+\mathbf{a}_2\), \(2\mathbf{a}_2-\mathbf{a}_2\), \(\mathbf{a}_2-2\mathbf{a}_1\), and their opposites. To make \ref{FigFS} we took \(\mu=0\), \(t=1\), and \(t_{nnn}=2.5\).

\begin{figure}[htp]
	\centering
	\includegraphics[width=0.49\textwidth]{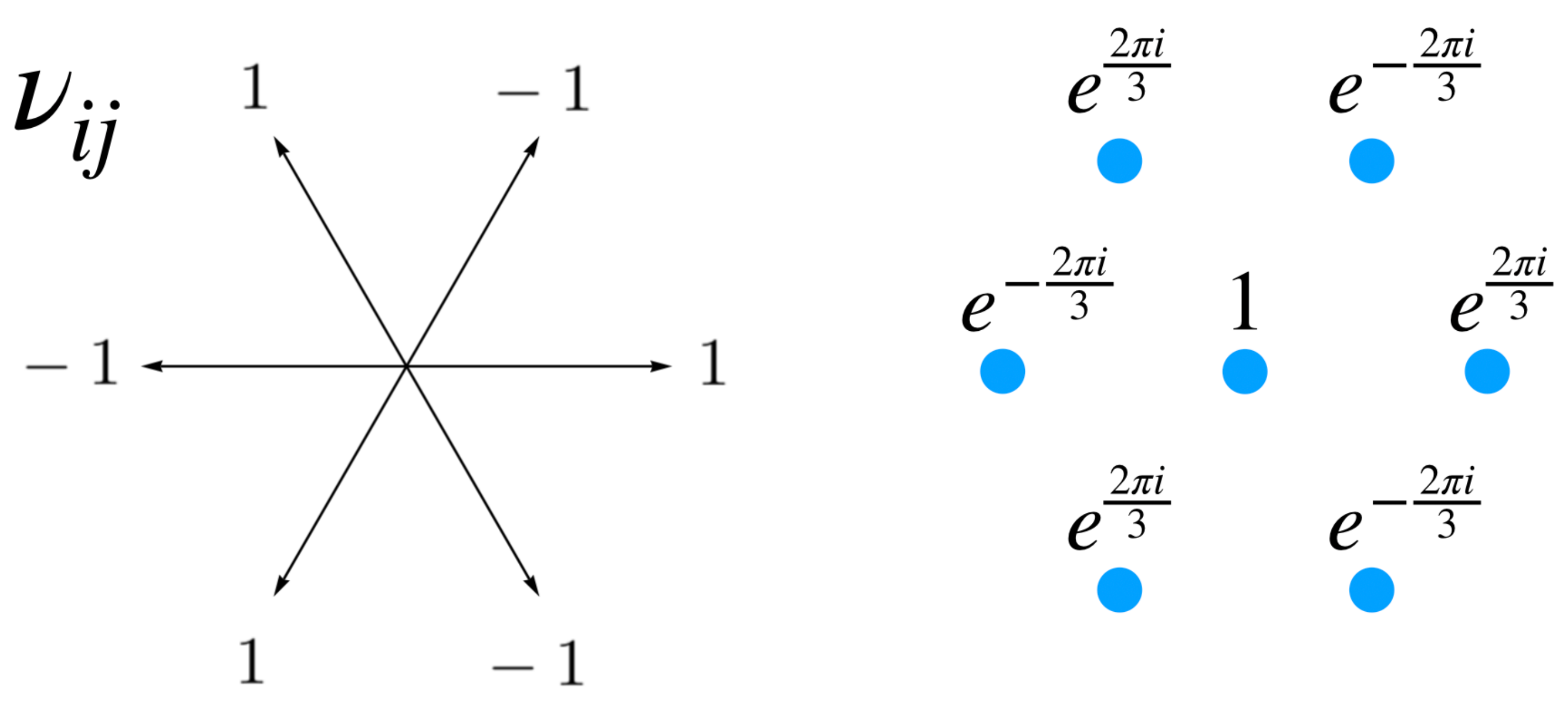}
	\caption{\(\nu_{ij}\) and relative phases of nearest neighbors in FF-type PDW\(_{K}\) with total momentum \(K\) in real space (opposite phases for PDW\(_{-K}\)).}
	\label{RealSpacePDWfig}
\end{figure}

The PDW Hamiltonian can also be included in the tight binding model. In particular, the spin-singlet (onsite) PDW\(_{-K}\) enters the Hamiltonian in real space on a lattice as (that can be obtained by Fourier transforming Eq. \ref{HbdgPDW}):
\[H_{PDW_-}=\sum_{j\alpha\beta}e^{-i\mathbf{K}\cdot\mathbf{R}_j} \Delta^{(s)}_{-K}i\sigma^y_{\alpha\beta}d^\dagger_{\mathbf{R}_j\alpha}d^\dagger_{\mathbf{R}_j\beta}+h.c.\label{PDWrealSpaceSinglet}\]
while the spin-triplet PDW\(_{-K}\), to leading order in lattice harmonics, enters as
\[H_{PDW_-}=\sum_{\langle ij\rangle \alpha\beta}e^{-i\mathbf{K}\cdot\mathbf{R}_j}\nu_{ij}\Delta^{(t)}_{-K}(\sigma^k i\sigma^y)_{\alpha\beta}d^\dagger_{\mathbf{R}_i\alpha}d^\dagger_{\mathbf{R}_j\beta}+h.c.\label{PDWrealSpaceTriplet}\]
with \(k=x,y,z\). The triplet order involves nearest neighbor terms with \(\nu_{ij}=\pm1\) if \(\mathbf{R}_j-\mathbf{R}_i\) points along \(\pm K\) (see Fig. \ref{RealSpacePDWfig}). Note that for nearest neighbors, \(e^{-i\mathbf{K}\cdot(\mathbf{R}_j-\mathbf{R}_i)}=e^{\pm 2\pi i/3}\) for \(\mathbf{R}_j-\mathbf{R}_i\) pointing along \(\pm K\), for any choice of the three equivalent \(\mathbf{K}\). The three-fold rotational symmetry therefore remains unbroken. Moreover, comparing phases of the order parameter going around a triangular loop, a total phase of \(\pm2\pi\) is accumulated, which indicates a loop supercurrent (see Fig. \ref{LoopFig}). Equivalently, we can think of three currents propagating along the three equivalent \(-\mathbf{K}\) directions, as one would expect since the Cooper pairs have a total momentum of \(-\mathbf{K}\). If PDW\(_{K}\) is also present, the currents cancel by superposition.

\section{Mean Field Solution of Linearized Gap Equation for Uniform SC}\label{AppendixB}

For completeness, we include here the reduced linearized gap equation for the uniform SC channels, which were previously obtained in \cite{Shaffer20}. Defining
\begin{eqnarray}\nonumber\label{gs2}
&\hat{g}_1^{(\mu)}=\bar{\Pi}_{\Gamma\Gamma}^{(\mu)} g_1^{(\mu)},\quad \hat{g}_{4\eta}^{(\mu)}=\bar{\Pi}_{\eta,-\eta}^{(\mu)}g_4^{(\mu)}, \quad \hat{g}_{23}^{(\mu)}=\bar{\Pi}_{K, -K}^{(\mu)} g_{23}^{(\mu)}
\end{eqnarray}
where $ g_{23}^{s}=\frac{g_2+g_3}{2}$, $g_{23}^{t}=\frac{g_2-g_3}{2}$,
the reduced linearized gap equation for the uniform SC channels read
\[ \label{Eq:BCSGapSol}
\left(\begin{array}{c}
\Delta^{(\mu)}_{\Gamma\Gamma}\\
\Delta^{(\mu)}_{K, -K}
\end{array}\right)=\left(\begin{array}{ccc}
\hat{g}^{(\mu)}_1 & 2\hat{g}^{(\mu)}_{4K}\\
\hat{g}^{(\mu)}_{4\Gamma} & 2\hat{g}^{(\mu)}_{23}
\end{array}\right)\left(\begin{array}{c}
\Delta^{(\mu)}_{\Gamma\Gamma}\\
\Delta^{(\mu)}_{K,-K}
\end{array}\right)\]
The matrix in the equation has eigenvalues
\[\gamma^{(\mu\pm)}=\frac{1}{2} \left(\hat{g}_1^{(\mu)}+2\hat{g}^{(\mu)}_{23}\pm \sqrt{(\hat{g}_1^{(\mu)}-2\hat{g}^{(\mu)}_{23})^2+8\hat{g}^{(\mu)}_{4\Gamma}\hat{g}^{(\mu)}_{4K}} \right)\]
in the SC channel. As in the PDW channels, the minus eigenvalue is always subleading and only the plus solution is relevant. The eigenvectors are
\[\left(\begin{array}{c}
\Delta^{(\mu+)}_{\Gamma\Gamma}\\
\Delta^{(\mu+)}_{K-K}
\end{array}\right)\propto\left(\begin{array}{c}
 \gamma^{(\mu)}_{+} - 2 \hat{g}^{(\mu)}_{23} \\
\hat{g}^{(\mu)}_{4\Gamma}
\end{array}\right)\]
In order for the PDW instability to be the leading one, one must have \(\kappa^{(\mu+)}>\gamma^{(\mu +)}\) (see Eq. \ref{Eq:Kappas}).

\section{Microscopic Derivation of the Ginzburg-Landau Free Energy}\label{AppendixC}

Here we outline the standard microscopic derivation of the free energy in Eq. \ref{F4PDW}. We use the Matsubara formalism with fermionic Matsubara frequencies \(\omega_n=\pi(2n+1)T\), and begin by carrying out the Hubbard-Stratonovich transformation of the action with Hamiltonian \(H+H_{int}\) (given by Eqs. \ref{H0} and \ref{HV}), which introduces bosonic fields \(\widehat{\Delta}_{\eta\zeta}\) and its conjugate \(\widehat{\Delta}^*_{\eta\zeta}\) and yields the action
\begin{align}\label{HubStratS2}
S&=\sum_{\substack{\mathbf{p,k}\\ \eta\zeta\eta'\zeta'}}\left[\widehat{\Delta}^*_{\eta'\zeta'}(\mathbf{k})\right]_{\alpha'\beta'}\left[V^{-1}(\mathbf{p;k})\right]^{\eta'\zeta';\alpha'\beta'}_{\eta\zeta;\alpha\beta}\left[\widehat{\Delta}_{\eta\zeta}(\mathbf{p})\right]_{\alpha\beta}+\nonumber\\
&+\frac{T}{2}\sum_{\substack{n,\mathbf{p}\\ \eta\zeta\alpha\beta}}\bar{\Psi}_{\mathbf{p}\eta\alpha}\left[-i\omega_n+\mathcal{H}^{(BdG)}_{\eta\zeta}(\mathbf{p})\right]_{\alpha\beta} \Psi_{\mathbf{p}\zeta\beta} \equiv\nonumber\\
&\equiv\frac{T}{2}\sum_{\substack{n,\mathbf{p}\\ \eta\zeta\alpha\beta}}\bar{\Psi}_{\mathbf{p}\eta\alpha}\left[-i\omega_n+\mathcal{H}^{(BdG)}_{\eta\zeta}(\mathbf{p})\right]_{\alpha\beta} \Psi_{\mathbf{p}\zeta\beta} +\beta H_{\Delta^2}
\end{align}
where we defined the Nambu spinors \(\Psi_{\mathbf{p}\eta\alpha}=(d_{\mathbf{p}\eta\alpha},\bar{d}_{-\mathbf{p}\eta\alpha})^T\) and the Bogolyubov-de Gennes (BdG) Hamiltonian
\[\mathcal{H}^{(BdG)}_{\eta\zeta}(\mathbf{p})=\left(\begin{array}{cc}
\mathcal{H}_{\eta}(\mathbf{p})\delta_{\eta\zeta} & \widehat{\Delta}_{\eta\zeta}(\mathbf{p}) \\
\widehat{\Delta}^\dagger_{\zeta\eta}(\mathbf{p}) & -\mathcal{H}^T_\zeta(-\mathbf{p})\delta_{\eta\zeta}
\end{array}\right)\]
where
\[\mathcal{H}_\eta(\mathbf{p})=\epsilon_\eta(\mathbf{p})+\mathbf{g}_\eta(\mathbf{p})\cdot\boldsymbol{\sigma}\label{H1bMF}\]
is the single-body normal state Hamiltonian with spin orbit coupling \(\mathbf{g}_\eta(\mathbf{p})\) that from now on we set to zero.

In the next step we integrate out the fermionic degrees of freedom \(d_{\mathbf{p}\eta\alpha}\) inside the path integral, which transforms the action into (dropping some constants and making the units of the logarithm dimensionless)
\[\label{Sdelta}S[\widehat{\Delta},\widehat{\Delta}^*]=-\sum_{n,\mathbf{p}}\log\det\left[\beta\left(-i\omega_n+\mathcal{H}^{(BdG)}(\mathbf{p})\right)\right]+\beta H_{\widehat{\Delta}^2}\] or equivalently the free energy \(\mathcal{F}=TS\):
\[\label{Fdelta}
\mathcal{F}[\widehat{\Delta},\widehat{\Delta}^*]=-T\sum_{n,\mathbf{p}}\text{Tr}\left[\log\beta\mathcal{G}^{-1}(i\omega_n,\mathbf{p})\right]+H_{\Delta^2}\]
where we defined the Nambu-Gor'kov Green's function as
\begin{align}
\mathcal{G}_{\eta\zeta}(i\omega,\mathbf{p})&=\left(i\omega-\mathcal{H}^{(BdG)}(\mathbf{p})\right)^{-1}_{\eta\zeta}=\nonumber\\
&=\left(\begin{array}{cc}
G_{\eta\zeta}(i\omega,\mathbf{p}) & F_{\eta\zeta}(i\omega,\mathbf{p}) \\
F_{\zeta\eta}^\dagger(i\omega,\mathbf{p}) & -G^T_{\eta\zeta}(-i\omega,-\mathbf{p})
\end{array}\right)
\end{align}
It is convenient to define the normal state Nambu-Gor'kov Green's function via
\[\mathcal{G}_0^{-1}(i\omega,\mathbf{p})=\mathcal{G}^{-1}(i\omega,\mathbf{p})-\widecheck{\Delta}(\mathbf{p})\]
where
\[\widecheck{\Delta}_{\eta\zeta}(\mathbf{p})=\left(\begin{array}{cc}
0 & \widehat{\Delta}_{\eta\zeta}(\mathbf{p}) \\
\widehat{\Delta}^\dagger_{\zeta\eta}(\mathbf{p}) & 0
\end{array}\right)\]
(i.e. \(\mathcal{G}_0^{-1}(i\omega,\mathbf{p})\) is the Nambu-Gor'kov Green's function in the limit when the gap functions vanish). When the gap function is small, we can then expand
\[\text{Tr}\left[\log\beta\mathcal{G}^{-1}\right]=\text{Tr}\left[\log\beta\mathcal{G}_0^{-1}\right]+\sum_j\frac{(-1)^{j+1}}{j}\text{Tr}\left[(\mathcal{G}_0\widecheck{\Delta})^j\right]\]
The odd \(j\) terms vanish (as they must since the free energy is a real scalar), which gives
\begin{widetext}
\[\mathcal{F}[\widehat{\Delta},\widehat{\Delta}^*]=T\sum_{n\mathbf{p} j} \frac{1}{2j}\text{Tr}\left[\left(\widehat{\Delta}^\dagger(\mathbf{p})G^{(0)}(i\omega_n,\mathbf{p}) \widehat{\Delta}(\mathbf{p})G^{(0,h)}(i\omega_n,\mathbf{p})\right)^j\right]+H_{\widehat{\Delta}^2}\label{Fexpansion}\]
where \(j\) is again summed over all positive integers and we defined the hole normal state Green's function \(G^{(0,h)}_{\eta\zeta}(i\omega,\mathbf{p})=-G_{\zeta\eta}^T(-i\omega,-\mathbf{p})\), and the trace is over pocket and spin indices. For example,
\begin{align}
&\text{Tr}\left[\widehat{\Delta}^\dagger(\mathbf{p})G^{(0)}(i\omega_n,\mathbf{p}) \widehat{\Delta}(\mathbf{p})G^{(0,h)}(i\omega_n,\mathbf{p})\right]=\sum_{\eta\zeta\eta'}\text{Tr}\left[\widehat{\Delta}^\dagger_{\eta\zeta}(\mathbf{p})G^{(0)}_\zeta(i\omega_n,\mathbf{p}) \widehat{\Delta}_{\zeta\eta'}(\mathbf{p})G_{\eta'}^{(0,h)}(i\omega_n,\mathbf{p})\right]=\nonumber\\
&=\sum_{\substack{\eta\zeta\eta'\\ \alpha\beta\gamma\delta}}\left[\widehat{\Delta}^\dagger_{\eta\zeta}(\mathbf{p})\right]_{\alpha\beta}\left[G^{(0)}_\zeta(i\omega_n,\mathbf{p})\right]_{\beta\gamma} \left[\widehat{\Delta}_{\zeta\eta'}(\mathbf{p})\right]_{\gamma\delta}\left[G_{\eta'}^{(0,h)}(i\omega_n,\mathbf{p})\right]_{\delta\alpha}
\end{align}
with additional spin indices \(\gamma\) and \(\delta\). The terms are easiest to compute diagrammatically, as shown in Fig. \ref{FexansionFig}.

\begin{figure*}[htp]
	\centering
	\includegraphics[width=0.9\textwidth]{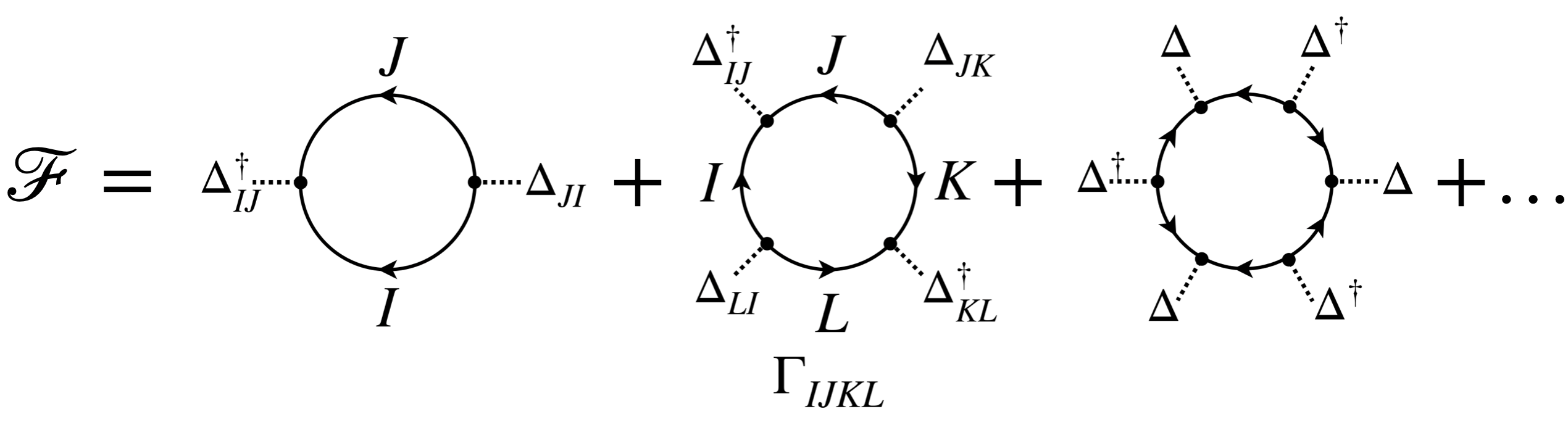}
	\caption{\label{FexansionFig}Diagrammatic expansion of the free energy in powers of the gap function.}
\end{figure*}

Minimizing the free energy with respect to \(\widehat{\Delta}^\dagger_{\eta\zeta}\) with only the \(j=1\) term in Eq. \ref{Fexpansion} yields the linearized gap equation. Assuming the Green's functions are diagonal in the spin indices, we can simplify the fourth order term (omitting \(\omega\) and \(\mathbf{p}\) that are all equal and an implicit sum over repeated indices):
\[\label{F4pockets}
\mathcal{F}^{(4)}=\beta_{\eta\zeta;\alpha\beta}^{\eta'\zeta';\alpha'\beta'}\left[\widehat{\Delta}^\dagger_{\eta\zeta}\right]_{\alpha\beta} \left[\widehat{\Delta}_{\zeta\eta'}\right]_{\beta\alpha'} \left[\widehat{\Delta}^\dagger_{\eta'\zeta'}\right]_{\alpha'\beta'} \left[\widehat{\Delta}_{\zeta'\eta}\right]_{\beta'\alpha}\]
where we define (assuming the gap functions only depend on the momentum directions and not their magnitudes)
\[\beta_{\eta\zeta;\alpha\beta}^{\eta'\zeta';\alpha'\beta'}(\theta)=\frac{T}{4}\sum_{n,|\mathbf{p}|} G^{(0)}_{\zeta\beta}G^{(0,h)}_{\eta'\alpha'}G^{(0)}_{\zeta'\beta'}G^{(0,h)}_{\eta\alpha}\]
For completeness, we mention that the Matsubara sum can be evaluated analytically using partial fractions and yields
\[\beta_{IJ}^{KL}=\sum_{|\mathbf{p}|}\frac{1}{\Sigma_{IJ}\Sigma_{KL}}\left[\frac{\tanh\frac{\beta\epsilon_{K}}{2}-\tanh\frac{\beta\epsilon_{I}}{2}}{\delta\epsilon_{IK}}+\frac{\tanh\frac{\beta\epsilon_{I}}{2}+\tanh\frac{\beta\epsilon_{L}}{2}}{\Sigma_{IL}}+\frac{\tanh\frac{\beta\epsilon_{J}}{2}+\tanh\frac{\beta\epsilon_{K}}{2}}{\Sigma_{JK}}+\frac{\tanh\frac{\beta\epsilon_{L}}{2}-\tanh\frac{\beta\epsilon_{J}}{2}}{\delta\varepsilon_{JL}}\right]\]
where the multi-indices include pocket and pseudospin indices (e.g. \(I=(\eta,\alpha)\)). Note that  \(I, K\) also carry the momentum \(\mathbf{p}\) while \(J, L\) carry momentum \(-\mathbf{p}\). We also defined \(\Sigma_{IJ}=\epsilon_I+\epsilon_J\) and \(\delta\epsilon_{IJ}=\epsilon_I-\epsilon_{J}\). In general the integral over \(|\mathbf{p}|\) cannot be evaluated analytically, but in the special case of equal DOS (this assumption can relaxed) with the assumption that \(\epsilon_I=\epsilon+\widehat{\Delta}_I\) where \(\widehat{\Delta}_I\) are only functions of the direction of \(\mathbf{p}\), the integral can be evaluated by performing a contour integration over momentum first, and then calculating the Matsubara sum. We obtain
\[\beta_{IJ}^{KL}=-\frac{N}{4\delta\epsilon_{IK}\delta\epsilon_{JL}}\text{Re}\left[\psi\left(\frac{1}{2}+\frac{i\delta\epsilon_{IJ}}{4\pi T}\right)-\psi\left(\frac{1}{2}+\frac{i\delta\epsilon_{IL}}{4\pi T}\right)+\psi\left(\frac{1}{2}+\frac{i\delta\epsilon_{KL}}{4\pi T}\right)-\psi\left(\frac{1}{2}+\frac{i\delta\epsilon_{JK}}{4\pi T}\right)\right]\label{GammaDigamma}\]
\end{widetext}
where \(\psi\) is the digamma function. In the limit of perfect nesting (i.e. all \(\delta\epsilon\rightarrow0\)), this simply evaluates to
\[\beta_{IJ}^{KL}\rightarrow \beta_0=\frac{7\zeta(3)N}{32\pi^2T^2}\]
where \(\zeta(3)\approx1.202\) is the Riemann zeta function (this can be found by setting \(\delta\epsilon=0\) before doing the Matsubara sum). Eq. \ref{F4pockets} can then be written more compactly as
\[\mathcal{F}^{(4)}=\beta_0\int\text{Tr}\left[\widehat{\Delta}^\dagger\widehat{\Delta}\widehat{\Delta}^\dagger\widehat{\Delta}\right]\frac{d\theta}{2\pi}\]
with the trace over both spin and pocket indices. Expressions in Eq. (\ref{betas}) follow after performing the elementary integrals. Note that the coefficients are more generally model dependent.

In the triplet channels, we have according to Eq. (\ref{GapExpanPDW})
\[\widehat{\Delta}_{\eta\zeta}^{(t)}(\mathbf{p})=\hat{\mathbf{d}}_{\eta+\zeta}\cdot \widehat{\boldsymbol{\Sigma}}^{(t)}_{\eta\zeta}(\theta)\Delta_{\eta\zeta}^{(t)}\]
with \(\widehat{\boldsymbol{\Sigma}}^{(t)}_{\eta\zeta}\propto \boldsymbol{\sigma}i\sigma^y\). The traces over Pauli matrices therefore reduce to scalars built out of the \(\mathbf{d}\) vectors, which can be evaluated using the identity
\[\left(\mathbf{d}_\eta\cdot\boldsymbol{\sigma}\right)\left(\mathbf{d}_\zeta\cdot\boldsymbol{\sigma}\right)=\left(\mathbf{d}_\eta\cdot\mathbf{d}_\zeta\right)+i\left(\mathbf{d}_\eta\times\mathbf{d}_\zeta\right)\cdot\boldsymbol{\sigma}\]
For the fourth order terms we therefore get terms of the form
\begin{align}
&\text{Tr}\left[\left(\mathbf{d}_\eta\cdot\boldsymbol{\sigma}\right)\left(\mathbf{d}^*_{\eta'}\cdot\boldsymbol{\sigma}\right)\left(\mathbf{d}_\zeta\cdot\boldsymbol{\sigma}\right)\left(\mathbf{d}^*_{\zeta'}\cdot\boldsymbol{\sigma}\right)\right]=\\
&=2\left(\mathbf{d}_\eta\cdot\mathbf{d}_{\eta'}^*\right)\left(\mathbf{d}_\zeta\cdot\mathbf{d}_{\zeta'}^*\right)-2\left(\mathbf{d}_\eta\times\mathbf{d}_{\eta'}^*\right)\cdot\left(\mathbf{d}_\zeta\times\mathbf{d}_{\zeta'}^*\right)\nonumber
\end{align}
For the \(\beta_4\) term in the free energy Eq. (\ref{F4PDW}), for example, we have \(\eta=\eta'=K\) and \(\zeta=\zeta'=-K\), so the trace reduces to \(2|\mathbf{d}_K|^2|\mathbf{d}_{-K}|^2=2\), since we assume a unit \(\mathbf{d}\) vector and unitary pairing is energetically favored (due to the negative sign in the last term). As a result, the form of the free energy for singlet and triplet PDW channels ends up having the same form, as claimed in the main text. Note that \(\mathbf{d}_K\) and \(\mathbf{d}_{-K}\)  are therefore independent of each other at the fourth order in the free energy in the absence of uniform SC.

Interestingly, this is no longer the case at sixth order of the free energy, given in Eq. (\ref{F6}). Again assuming unitary pairing, the only term that couples \(\mathbf{d}_K\) and \(\mathbf{d}_{-K}\)  is \(\Gamma_3\), with the relevant trace being
\begin{align}
&\text{Tr}\left[\left[\left(\mathbf{d}_K\cdot\boldsymbol{\sigma}\right)\left(\mathbf{d}^*_{-K}\cdot\boldsymbol{\sigma}\right)\right]^3\right]=\\
&=2\left(\mathbf{d}_K\cdot\mathbf{d}^*_{-K}\right)^3-3\left(\mathbf{d}_K\cdot\mathbf{d}^*_{-K}\right)\left(\mathbf{d}_K\times\mathbf{d}^*_{-K}\right)^2\nonumber
\end{align}
which is minimized (maximized) when \(\mathbf{d}_K\) and \(\mathbf{d}_{-K}^*\) are aligned in the same (opposite) direction. We therefore assume them to be aligned in the main text.

\bibliographystyle{apsrev4-1}
\bibliography{PDW}

\end{document}